\documentclass[aps,groupedaddress,showpacs,floatfix]{revtex4}
\usepackage{amssymb}
\usepackage{amsmath}
\usepackage{graphicx}
\begin{document}

\title{Velocity distribution functions and intermittency in one-dimensional randomly forced Burgers turbulence}

\author{Victor Dotsenko}

\affiliation{Sorbonne Universit\'e, LPTMC, F-75005 Paris, France}

\affiliation{L.D.\ Landau Institute for Theoretical Physics,
 119334 Moscow, Russia}

\date{\today}

\begin{abstract}

The problem of one-dimensional randomly forced Burgers turbulence is
considered in terms of (1+1) directed polymers.
In the limit of strong turbulence (which corresponds to the zero temperature limit for
the directed polymer system) using the replica technique
a general explicit expression for the joint distribution function of two velocities
separated by a finite distance is derived. In particular,
it is shown that at length scales much smaller than the injection length
of the Burgers random force the moments of the velocity increment exhibit
typical strong intermittency behavior.

\end{abstract}

\pacs{
      05.20.-y  
      75.10.Nr  
      74.25.Qt  
      61.41.+e  
     }

\maketitle

\section{Introduction}

In the problem of one-dimensional randomly forced Burgers turbulence one studies the statistical properties
a velocity field $v(x,t)$ governed by the Burgers equation \cite{burgers_74}
\begin{equation}
   \label{1}
\partial_{t} v(x,t) + v(x,t)\partial_{x} v(x,t) =
\nu  \partial^{2}_{x} v(x,t) +  f(x,t)
\end{equation}
where the parameter $\nu$ is the viscosity and $f(x,t)$ is the Gaussian distributed random force
which is $\delta$-correlated in time and which is characterized by finite correlation length $R$ in space:
$\overline{f(x,t) f(x',t')} = u \delta(t-t') {\cal F}[(x-x')/R]$. Here ${\cal F}(x)$ is a smooth function decaying to zero fast
enough at large arguments and the parameter $u$ is the injected energy density. This problem is the subject of
active investigations for more that six decades (see e.g. \cite{Sinai,Bouch-Mez-Par,Khanin} and references there in).

In the framework of the celebrated Kolmogorov theory \cite{Kolmogorov1} one obtains the probability density function (PDF)
of the velocity increment $w = v(x_{0}+x,t)-v(x_{0},t)$, such that at distances much smaller than the length scale $R$
of the random stirring force $f$, one finds simple scaling for the moments $\langle w^{q}\rangle \sim x^{\zeta(q)}$  with $\zeta(q) = n/3$
(in particular one can prove that $\zeta(3) =1$). This prediction is based on the assumption that the statistical properties
of the velocity field is locally homogeneous,  so that the corresponding PDF of $w$ depends only on $r$
and the average rate of the energy dissipation. However, extensive studies during last decades
convincingly demonstrate that in fact the exponent $\zeta(q)$ significantly deviates from the Kolmogorov's law $q/3$.
The physical reason for that is  the so called intermittency phenomenon, namely,
formation of local coherent structures that drives a strong deviation from the mean fluctuation level
of the velocity field
\cite{Kolmogorov2,Obukhov,intermittency1,intermittency2,intermittency3,intermittency4,intermittency5,intermittency6,intermittency7}

In the present paper using formal equivalence of the above Burger's problem, eq.(\ref{1}), with the model of one-dimensional
directed polymers in a random potential \cite{Bouch-Mez-Par} (see below) we are going to derive en explicit expression for the
joint  PDF $P(v, v')$ for two velocities separated by a distance $x$, as well as the corresponding PDF for the velocity increment
$w = v'-v$. In particular, at distances much smaller than the scale  of the stirring force, $x \ll R$
this  allows to demonstrate the typical intermittency behavior of the exponent $\zeta(q)$ (see Fig.1)


\vspace{5mm}

It is well known that the  Burgers problem, eq.(\ref{1}), is formally equivalent to the one of growing interfaces
in a random environment described by the Kardar-Parisi-Zhang (KPZ) equation \cite{KPZ,hh_zhang}.
Indeed, redefining
\begin{equation}
 \label{2}
v(x,t) = -\partial_{x} F(x,t)
\end{equation}
and $f(x,t) = -\partial_{x} V(x,t)$, and integrating once eq.(\ref{1}),
one gets the KPZ equation for the interface profile $F(x,t)$,
\begin{equation}
   \label{3}
\partial_{t} F(x,t) = \frac{1}{2} \bigl(\partial_{x} F(x,t)\bigr)^{2}
+ \nu \partial^{2}_{x} F(x,t) +  V(x,t)
\end{equation}
where  $V(x,t)$ is a random potential.
On the other hand, let us consider one-dimensional directed polymers system which
is defined in terms of the Hamiltonian
\begin{equation}
   \label{4}
   H[\phi(\tau), V] = \int_{0}^{t} d\tau
   \Bigl\{\frac{1}{2} \bigl[\partial_\tau \phi(\tau)\bigr]^2
   + V[\phi(\tau),\tau]\Bigr\};
\end{equation}
where $\phi(\tau)$ is a scalar field defined
within an interval $0 \leq \tau \leq t$ and
$V(\phi,\tau)$ is the Gaussian distributed random potential with a zero mean, $\overline{V(\phi,\tau)}=0$,
and the correlation function
\begin{equation}
\label{5}
\overline{V(\phi,\tau)V(\phi',\tau')} = u \delta(\tau-\tau') U(\phi-\phi')
\end{equation}
Here the parameter $u$ defines the strength of the disorder and $U(\phi)$ is the spatial correlation
function characterized by the correlation length $R$. For simplicity we take
\begin{equation}
\label{6}
U(\phi) \; = \; \frac{1}{\sqrt{2\pi} \, R} \; \exp\Bigl\{-\frac{\phi^{2}}{2 R^{2}}\Bigr\}
\end{equation}
For a given realization of the random potential $ V[\phi,\tau]$
the partition function of this system  is defined as
\begin{equation}
\label{7}
Z(x,t) = \int_{\phi(0)=0}^{\phi(t)=x} {\cal D}\phi(\tau)
        \exp\bigl\{-\beta H[\phi(\tau), V]\bigr\} \; = \; \exp\bigl\{-\beta F(x,t)\bigr\}
\end{equation}
where $\beta$ is the inverse temperature,
$F(x,t)$ is the free energy  and the integration is taken over all
trajectories $\phi(\tau)$ with the boundary conditions at $\phi(\tau=0) = 0$ and $\phi(\tau=t) = x$.
One can easily show that the partition function $Z(x,t)$ defined above satisfy the linear differential equation
\begin{equation}
 \label{8}
\partial_{t} Z(x,t) \; = \;  \frac{1}{2\beta} \partial^{2}_{x} Z(x,t) \; - \; \beta  V(x,t) Z(x,t)
\end{equation}
Substituting here $Z(x,t) = \exp\bigl\{-\beta F(x,t)\bigr\}$, one easily finds that the free energy function $F(x,t)$
satisfy the KPZ equation (\ref{3}) with the viscosity parameter $\nu = \frac{1}{2\beta}$.
In other words, the original random force Burger's problem, eq.(\ref{1}) is formally equivalent the
directed polymer system, eqs.(\ref{4})-(\ref{7}), such that the the viscosity parameter $\nu$ in the Burger's equation
is proportional to the temperature in the directed polymer system, $\nu = \frac{1}{2} T$, and the velocity $v(x,t)$
in the Burger's equation is the negative spatial derivative of the free energy $F(x,t)$ of the directed polymers system.

The standard dimensionless parameter which characterizes the level
of turbulence of the velocity field in the Burgers problem is called the Reynolds number $Re$, and
it is defined as  the ratio of typical values of the inertial forces to viscous forces.
In the present notations
it can be defined as $Re = v_{0} R/\nu$,
where $v_{0}$ is the typical flow velocity at the characteristic linear dimension
which in the present case is the injection scale of the random force $R$.
Using dimensional arguments one easily finds that
\begin{equation}
 \label{9a}
v_{0} \; \sim \; \Bigl(\frac{u}{R^{2}}\Bigr)^{1/3}
\end{equation}
Indeed, according to eq.(\ref{2}) the dimension of the velocity $[v_{0}] = [F]/R$. On the other hand according to
eq.(\ref{4}), the dimension of the free energy is $[F] = [H] = t [V]$. Finally, according to eqs.(\ref{5})-(\ref{6}),
the dimension of the random potential is $[V] = \sqrt{u/(Rt)}$. Combining all that together one finds eq.(\ref{9a}).
Therefore, in terms of the directed polymers notations the Reynolds number of the Burgers turbulence problem reads
\begin{equation}
 \label{9}
Re \; = \; \frac{v_{0} R}{\nu} \; = \; 2\beta \, \bigl(u R\bigr)^{1/3}
\end{equation}
It is evident that an increasing Reynolds number indicates an increasing turbulence of flow and
the limit of strongly developed turbulence corresponds to $Re \to \infty$.
Thus the strong turbulence Burgers regime  corresponds to the zero-temperature limit
in the directed polymers system and it is this limit which will be studied in the present paper.

As the velocity in the Burgers problem is given by the spatial derivative of the free energy
of the directed polymer system, it can be expressed in terms of the difference of two free energies:
\begin{equation}
 \label{10}
v(x.t) \; = \; -\frac{\partial F(x,t)}{\partial x} \; = \; - \lim_{\epsilon\to 0} \frac{F(x+\epsilon, t) - F(x,t)}{\epsilon}
\end{equation}
In other words, one-point velocity statistics  is defined by the joint statistics of of {\it two} free energies.
Correspondingly, if we are going to study the joint statistical properties of two spatially separated velocities,
in terms of the free energies of the directed polymers we have to study the four-point spatial object.

In Section II we describe the general ideas and the main lines of the replica approach which will be used in the further
derivations of the  probability distribution
functions. In section III we describe the main points of the zero-temperature limit approach for the directed polymers with
finite correlation length of the random potential, eqs.(\ref{5})-(\ref{6}) (for details see \cite{zero-T}).
The zero temperature limit of the joint probability distribution function of free energies defined at four spatial points
is derived in Section IV. The explicit expression for the corresponding joint probability density function
of two velocities $v$ and $v'$ separated by a distance $x$ is derived in Section V, eqs.(\ref{110})-(\ref{112}).
In Section VI it will be shown that the PDF for the velocity increment $w = v-v'$ has the following form:
\begin{equation}
 \label{11}
P_{x}(w) \; = \; p_{0}(x/R) \delta\Bigl(w - v_{0}\frac{x}{R}\Bigr)
               \; + \; {\cal P}_{x/R} \bigl(w/v_{0}\bigr) \, \theta\Bigl(v_{0}\frac{x}{R} - w\Bigr)
\end{equation}
where $\theta(z)$ is the Heaviside step function, $v_{0} \, \propto \, \bigl(u/R^{2}\bigr)^{1/3}$ (see eq.(\ref{9a})),
\begin{equation}
 \label{12}
p_{0}(x/R) \; = \; \int_{-\infty}^{+\infty} \frac{ds}{\sqrt{2\pi}} \;
               \frac{\exp\Bigl\{-\frac{1}{2} s^{2}\Bigr\}}{
               \Biggl(1 \; + \; \frac{\zeta_{0}^{3/4} x}{R} \int_{0}^{+\infty} \frac{d\xi}{\sqrt{2\pi}}  \, \xi \,
                         \exp\Bigl\{-\frac{1}{2} (s+\xi)^{2}\Bigr\}\Biggr)}
\end{equation}
and
\begin{equation}
 \label{13}
{\cal P}_{x/R} \bigl(w/v_{0}\bigr) \; = \; \frac{\zeta_{0}^{3/2} x}{v_{0} R}
                \int_{-\infty}^{+\infty} \frac{ds}{\sqrt{2\pi}}
                \int_{0}^{\Delta(w/v_{0})} \frac{d\eta}{\sqrt{2\pi}}
        \frac{\exp\Bigl\{-\frac{1}{2} s^{2} -\frac{1}{2} \bigl(s-\Delta(w/v_{0})\bigr)^{2}\Bigr\}}{
               \Biggl(1  + \frac{\zeta_{0}^{3/4} x}{R} \int_{0}^{+\infty} \frac{d\xi}{\sqrt{2\pi}}  \, \xi \,
                         \exp\Bigl\{-\frac{1}{2} \bigl(\xi + \eta + s - \Delta(w/v_{0}) \bigr)^{2}\Bigr\}\Biggr)^{2}}
\end{equation}
Here $\zeta_{0} \sim 1 $ is a number (see Section III) and
\begin{equation}
 \label{14}
\Delta(w/v_{0}) \; = \; \zeta_{0}^{3/4} \Bigl(\frac{x}{R} \; - \; \frac{w}{v_{0}} \Bigr)
\end{equation}
The above formulas, eqs.(\ref{11})-(\ref{14}) constitute the central result of the present research.
The distribution function $P_{x}(w)$ has rather specific structure (see Section VI, Fig.4).
According to eq.(\ref{11}) for a given  distance $x$ the values of the velocity increment
$w$ are bounded from above:  $w \leq \frac{x}{R} v_{0}$, where  $v_{0} \propto \bigl(u/R^{2}\bigr)^{1/3}$
is the typical flow velocity at the injection scale $R$ of the random force of the strength $u$.
Moreover, at $w = \frac{x}{R} v_{0}$ the distribution function exhibits the $\delta$-function singularity
which means that at a given distance $x$ the difference of two velocities $w = v-v'$ has a {\it finite}
probability $p_{0}$, eq.(\ref{12}), to be equal to $\frac{x}{R} v_{0}$.

The above result allows to study the behavior of the moments of the velocity increment $\langle w^{q}\rangle$
at distances $x \ll R$.
Introducing the reduced distance parameter $r = \zeta_{0}^{3/4} x/R$ and the reduced velocity increment
$\omega = \zeta_{0}^{3/4} w/v_{0}$, in the limit $r \ll 1$ instead of eqs.(\ref{11})-(\ref{13}) we get:
\begin{equation}
 \label{15}
P_{r}(\omega) \; \simeq \; \Bigl(1-\frac{r}{\sqrt{\pi}}\Bigr) \delta(\omega - r)
               \; + \; \frac{r}{\sqrt{\pi}} (r-\omega) \exp\Bigl\{-\frac{1}{4}(r-\omega)^{2}\Bigr\} \; \theta(r-\omega)
\end{equation}
Then, for even moments of the reduced velocity increment we obtain:
\begin{equation}
\label{16}
  \langle \omega^{2n}\rangle \; \simeq \; r^{2n} \; + \; C(n) \, r
 \end{equation}
where $C(n) = 2^{2n+1} \Gamma(1+n)$. The above result can be analytically continued for arbitrary real values $q$
of the parameter $2n \to q$. Then, introducing the exponent $\zeta(q)$ as
$\langle \omega^{q}\rangle \; \simeq \; r^{\zeta(q)}$, according to eq.(\ref{16}) in the limit $r \ll 1$
we recover the typical strong intermittency behavior (see Fig.1):
\begin{equation}
 \label{17}
\zeta(q) \; \simeq \;
\left\{
        \begin{array}{ll}
           q \; , \; \; \mbox{for} \; q \; \leq \; 1 \, ;
       \\
       \\
           1 \; , \; \; \mbox{for} \; q \; > \; 1 \, .
        \end{array}
\right.
\end{equation}
\begin{figure}[h]
\begin{center}
   \includegraphics[width=9.0cm]{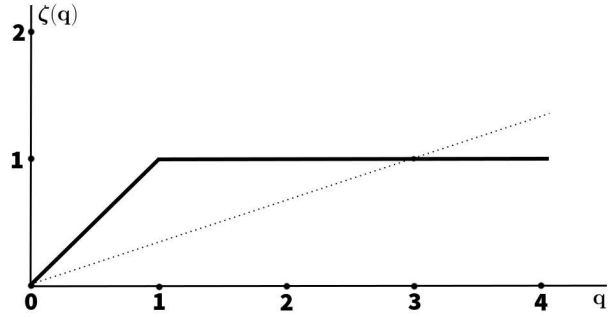}
\caption[]{Intermittency behavior of the exponent $\zeta(q)$, eq.(\ref{17}). The dashed line represents the Kolmogorov scaling
            $\zeta(q) = q/3$}
\end{center}
\label{figure1}
\end{figure}
It should be stressed that the above results, eqs.(\ref{11}) and (\ref{17}), are in remarkable agreement with the ones
obtained for the same system many years ago in \cite{Bouch-Mez-Par} in the framework of the  Gaussian variation method
(which formally should be valid only at high dimensions).

\section{Replica formalism}

In this section we are going to describe the general scheme of calculations of the statistical properties of the
Burgers velocity field $v(x,t)$, eq.(\ref{1}), in terms of the standard replica approach used for the directed polymers
model, eqs.(\ref{4})-(\ref{7}).
Using the relation between  $v(x,t)$ and the free energy $F(x,t)$ of the corresponding
directed polymer model, eq.(\ref{10}), for finite value of the parameter $\epsilon$
(which should be taken to zero in the final result) we have
\begin{equation}
 \label{18}
\exp\bigl\{\beta\epsilon v(x,t) \bigr\} \; = \; \exp\bigl\{ -\beta F(x+\epsilon, t) + \beta F(x, t) \bigr\}
                                        \; = \; Z(x+\epsilon, t) \cdot Z^{-1}(x, t)
\end{equation}
Taking the integer power $N$ of both sides of the above relation and averaging over the disorder
(which in what follows  will be denoted by the overline, $\overline{(...)}$) we gets
\begin{equation}
 \label{19}
\int dv \, P_{x,\epsilon,t}(v) \, \exp\bigl\{\beta N\epsilon \, v\bigr\} \; = \; \overline{Z^{N}(x+\epsilon, t) \cdot Z^{-N}(x, t)}
\end{equation}
where $P_{x,\epsilon,t}(v)$ is the PDF of the velocity $v$. Formally, the above relation can be represented as
follows,
\begin{equation}
 \label{20}
\int dv \, P_{x,\epsilon,t}(v) \, \exp\bigl\{\beta N\epsilon \, v\bigr\} \; = \; \lim_{M\to 0} \, Z(M,N,x,\epsilon,t)
\end{equation}
where
\begin{equation}
 \label{21}
Z(M,N,x,\epsilon,t) \; \equiv \; \overline{Z^{N}(x+\epsilon, t) \cdot Z^{M-N}(x, t)}
\end{equation}
is the {\it two-point} replica partition function.

General scheme of calculations of the
velocity PDF defined by the relation (\ref{20}) consists of several steps. First, for a given (finite) $\epsilon$
and {\it integers} $M$ and $N$, such that $M > N$, one has to compute the replica partition function
$Z(M,N,x,\epsilon,t)$ as an analytic function of the parameters $M$ and $N$. Next, this function should be analytically continued
for arbitrary complex values of $M$ and $N$, and the limits $M \to 0$ as well as $t\to\infty$ have to be taken.
Next, to take the limit $\epsilon\to 0$, one introduces the parameter $s = \beta\epsilon N$ which
has to be kept finite (this implies that together with the limit $\epsilon\to 0$ one simultaneously takes the limit
$N\to\infty$). Thus, after performing these manipulations (provided all the above limits exist) the relation (\ref{20})
turns into bilateral  Laplace transform for the velocity PDF $P_{*}(v) = \lim_{\epsilon\to 0}\lim_{t\to\infty} P_{x,\epsilon,t}(v)$
(which for finite values of $x$ in the limit $t\to\infty$ should be $x$-independent):
\begin{equation}
 \label{22}
\int dv \, P_{*}(v) \, \exp\{s \, v\} \; = \; Z_{*}(s)
\end{equation}
where
\begin{equation}
  \label{23}
Z_{*}(s) \; = \; \lim_{\epsilon\to 0}\lim_{t\to\infty}\lim_{M\to 0} \; Z\Bigl(M,\frac{s}{\beta\epsilon},x,\epsilon,t\Bigr)
\end{equation}
In this way the PDF $P_{*}(v)$ could be recovered by the inverse Laplace transform.

Note that the above type of program
has been already successfully implemented for the derivation of the Burgers two-point velocity PDF in the toy (Gaussian) Larking model of
random directed polymers \cite{burgulence}.

In this paper we are going to derive joint PDF of two velocities $v = v(x/2,t)$ and $v'=v(-x/2,t)$ at two points separated by a {\it finite }
distance $x$. In this case straightforward generalization of the above replica scheme would require computation of the
four-point replica partition function:
\begin{equation}
 \label{24}
\int dv dv' \, P_{x,\epsilon,t}(v, v') \, \exp\bigl\{\beta N_{1}\epsilon \, v + \beta N_{2}\epsilon \, v'\bigr\}  =
      \lim_{M\to 0} \, \overline{Z^{N_{1}}(x/2,t) \, Z^{M-N_{1}}(x/2-\epsilon,t) \, Z^{N_{2}}(-x/2+\epsilon,t) \, Z^{M-N_{2}}(-x/2,t) }
\end{equation}
Technically, direct recovery (using the inverse Laplace transformation) of the two-velocity PDF using the above relation
turns out to be rather involved task which still remains to be done. On the other hand, the experience shows that sometimes
to compute a complicated quantity, first one just has to compute a more general object. In the present case, instead
of two velocities PDF let us consider the joint distribution function of {\it three} free energy differences.
Namely, for a given four spatial points, $-x/2, \; -x/2+\epsilon, \; x/2-\epsilon$ and $x/2$, let us define
\begin{eqnarray}
 \nonumber
f_{1} &=& F(x/2,t) \; - \; F(-x/2,t)
\\
\label{25}
f_{2} &=& F(x/2-\epsilon,t) \; - \; F(-x/2,t)
\\
\nonumber
f_{3} &=& F(-x/2+\epsilon,t) \; - \; F(-x/2,t)
\end{eqnarray}
In terms of the partition functions the above relations can be represented as follows:
\begin{eqnarray}
 \nonumber
Z(x/2,t) \, Z^{-1}(-x/2,t) &=& \exp\{-\beta f_{1}\}
\\
\nonumber
\\
\label{26}
Z(x/2-\epsilon,t) \, Z^{-1}(-x/2,t) &=& \exp\{-\beta f_{2}\}
\\
\nonumber
\\
\nonumber
Z(-x/2+\epsilon,t) \, Z^{-1}(-x/2,t) &=& \exp\{-\beta f_{3}\}
\end{eqnarray}
As the further considerations will be done in the zero temperature limit (which correspond to the limit of
large Reynolds number, eq.(\ref{9})), it turns out that
the simplest way to derive the joint PDF  $P_{x,\epsilon,t}\bigl(f_{1}, f_{2},f_{3}\bigr)$ is to use the
generating function approach. Namely, let us introduce the probability function
\begin{equation}
 \label{27}
W_{x,\epsilon,t}\bigl(f_{1}, f_{2},f_{3}\bigr) \; = \;
                             \int_{-\infty}^{f_{1}}df_{1}' \int_{-\infty}^{f_{2}}df_{2}' \int_{-\infty}^{f_{3}}df_{3}' \;
                             P_{x,\epsilon,t}\bigl(f_{1}', f_{2}',f_{3}'\bigr)
\end{equation}
One can easily see that in the zero temperature limit this function can be represented in a form of the series:
\begin{eqnarray}
 \nonumber
W_{x,\epsilon,t}\bigl(f_{1}, f_{2},f_{3}\bigr) &=& -\lim_{\beta\to\infty}
        \sum_{N_{1}=1}^{\infty} \frac{(-1)^{N_{1}}}{N_{1}!}
        \sum_{N_{2}=1}^{\infty} \frac{(-1)^{N_{2}}}{N_{2}!}
        \sum_{N_{3}=1}^{\infty} \frac{(-1)^{N_{3}}}{N_{3}!}
\exp\bigl\{\beta N_{1} f_{1}+\beta N_{2} f_{2}+\beta N_{3} f_{3}\bigr\}
\times
\\
\label{28}
\\
\nonumber
&\times&
\overline{\Bigl[Z(x/2,t) Z^{-1}(-x/2,t) \Bigr]^{N_{1}}
          \Bigl[Z(x/2-\epsilon,t) Z^{-1}(-x/2,t) \Bigr]^{N_{2}}
          \Bigl[Z(-x/2+\epsilon,t) Z^{-1}(-x/2,t)  \Bigr]^{N_{3}}      }
\end{eqnarray}
Indeed, substituting here eq.(\ref{26}) we gets
\begin{eqnarray}
 \nonumber
W_{x,\epsilon,t}\bigl(f_{1}, f_{2},f_{3}\bigr) &=& -\lim_{\beta\to\infty}
       \int_{-\infty}^{+\infty}df_{1}' \int_{-\infty}^{+\infty}df_{2}' \int_{-\infty}^{+\infty}df_{3}' \;
                             P_{x,\epsilon,t}\bigl(f_{1}', f_{2}',f_{3}'\bigr) \;
\Biggl[ \sum_{N_{1}=1}^{\infty} \frac{(-1)^{N_{1}}}{N_{1}!} \exp\bigl\{\beta(f_{1}-f_{1}')N_{1}\bigr\} \Biggr]
\times
\\
\nonumber
\\
\nonumber
&\times&
\Biggl[ \sum_{N_{2}=1}^{\infty} \frac{(-1)^{N_{2}}}{N_{2}!} \exp\bigl\{\beta(f_{2}-f_{2}')N_{2}\bigr\} \Biggr]
\Biggl[ \sum_{N_{3}=1}^{\infty} \frac{(-1)^{N_{3}}}{N_{3}!} \exp\bigl\{\beta(f_{3}-f_{3}')N_{3}\bigr\} \Biggr]
\\
\nonumber
\\
\nonumber
\\
\nonumber
&=& -\lim_{\beta\to\infty} \int_{-\infty}^{+\infty}df_{1}' \int_{-\infty}^{+\infty}df_{2}' \int_{-\infty}^{+\infty}df_{3}' \;
                             P_{x,\epsilon,t}\bigl(f_{1}', f_{2}',f_{3}'\bigr) \;
\Biggl[ \exp\Bigl\{-\exp\bigl[\beta(f_{1}-f_{1}')\bigr] \Bigr\} \, - \, 1 \Biggr]
\times
\\
\nonumber
\\
\nonumber
&\times&
\Biggl[ \exp\Bigl\{-\exp\bigl[\beta(f_{2}-f_{2}')\bigr] \Bigr\} \, - \, 1 \Biggr]
\Biggl[ \exp\Bigl\{-\exp\bigl[\beta(f_{3}-f_{3}')\bigr] \Bigr\} \, - \, 1 \Biggr]
\\
\nonumber
\\
\nonumber
\\
\label{29}
&=& \int_{-\infty}^{+\infty}df_{1}' \int_{-\infty}^{+\infty}df_{2}' \int_{-\infty}^{+\infty}df_{3}' \;
                             P_{x,\epsilon,t}\bigl(f_{1}', f_{2}',f_{3}'\bigr) \;
\theta\bigl(f_{1}-f_{1}'\bigr) \, \theta\bigl(f_{2}-f_{2}'\bigr) \, \theta\bigl(f_{3}-f_{3}'\bigr)
\end{eqnarray}
which coincides with the definition (\ref{27}).

Thus, according to eq.(\ref{28}), in terms of the replica technique the probability function, eq.(\ref{27}),
can be represented as:
\begin{equation}
 \label{30}
W_{x,\epsilon,t}\bigl(f_{1}, f_{2},f_{3}\bigr) = -\lim_{\beta\to\infty} \lim_{M\to 0}
        \sum_{N_{1},N_{2},N_{3}=1}^{\infty} \frac{(-1)^{N_{1}+N_{2}+N_{3}}}{N_{1}!N_{2}!N_{3}!}
        \exp\bigl\{\beta N_{1} f_{1}+\beta N_{2} f_{2}+\beta N_{3} f_{3}\bigr\} \;
   Z_{x,\epsilon,t} \bigl(M,N_{1},N_{2},N_{3} \bigr)
\end{equation}
where
\begin{equation}
 \label{31}
Z_{x,\epsilon,t} \bigl(M,N_{1},N_{2},N_{3} \bigr) \; = \;
     \overline{Z^{N_{1}}(x/2,t) \, Z^{N_{2}}(x/2-\epsilon,t) \, Z^{N_{3}}(-x/2+\epsilon,t) \, Z^{M-N_{1}-N_{2}-N_{3}}(-x/2,t) }
\end{equation}
Further program of calculations is in the following.
The above {\it four-point} replica partition function has to be calculated for an integer $M > N_{1}+N_{2}+N_{3}$ as an analytic
function of the parameter $M$. Then this function has to be analytically continued for arbitrary real values of $M$
and the limit $M\to 0$ has to be taken. Finally, after computing the series in eq.(\ref{30}) (in the limits $t\to\infty$ and
$\beta\to\infty$) according to the definition (\ref{27}) the corresponding PDF $P_{x,\epsilon}\bigl(f_{1}, f_{2},f_{3}\bigr)$
can be obtained as
\begin{equation}
 \label{32}
P_{x,\epsilon}\bigl(f_{1}, f_{2},f_{3}\bigr) \; = \;
   \frac{\partial^{3}}{\partial f_{1} \; \partial f_{2} \; \partial f_{3}} {\cal W}_{x,\epsilon}\bigl(f_{1}, f_{2},f_{3}\bigr)
\end{equation}
where
\begin{equation}
 \label{33}
{\cal W}_{x,\epsilon}\bigl(f_{1}, f_{2},f_{3}\bigr) \; \equiv \;
            \lim_{\beta\to \infty}  \lim_{t\to\infty} W_{x,\epsilon,t}\bigl(f_{1}, f_{2},f_{3}\bigr)
\end{equation}
According to the representation (\ref{10}) and the definitions (\ref{25}) 
the velocities $v \equiv v(x/2,t)$ and $v'\equiv v(-x/2,t)$ are defined as
\begin{eqnarray}
 \label{34}
v &=& - \lim_{\epsilon\to 0} \frac{f_{1} - f_{2}}{\epsilon}
\\
\label{35}
v' &=& - \lim_{\epsilon\to 0} \frac{f_{3}}{\epsilon}
\end{eqnarray}
Thus, the corresponding joint PDF of these two velocities, $P_{x}(v, v')$ can be obtained as
\begin{equation}
 \label{36}
P_{x}(v, v') \; = \; \lim_{\epsilon\to 0} \Biggl[\epsilon^{2} \int_{-\infty}^{+\infty} df_{2} \; 
              P_{x,\epsilon}\bigl(f_{2}-\epsilon v,\,  f_{2},\, -\epsilon v'\bigr) \Biggr]
\end{equation}
The above general program of computations  will be implemented in the further sections.

\section{Zero temperature limit}

To compute the replica partition function, eq.(\ref{31}), let us consider more general object:
\begin{equation}
 \label{37}
\Psi\bigl(x_{1}, x_{2}, ..., x_{M}; \, t\bigr) \; = \; \overline{\Biggl(\prod_{a=1}^{M} Z(x_{a}, t) \Biggr)}
\end{equation}
Substituting here eqs.(\ref{7}) and (\ref{4}), and performing simple Gaussian averaging (using eq.(\ref{5}))
we get
\begin{equation}
 \label{38}
\Psi\bigl(x_{1}, x_{2}, ..., x_{M}; \, t\bigr) \; = \;
     \prod_{a=1}^{M} \Biggl[\int_{\phi_{a}(0)=0}^{\phi_{a}(t)=x_{a}} {\cal D}\phi_{a}(\tau)\Biggr]
         \exp\Bigl\{-\beta H_{M} \bigl[\phi_{1}(\tau), ..., \phi_{M}(\tau)\bigr] \Bigr\}
\end{equation}
where
\begin{equation}
 \label{39}
\beta H_{M} \bigl[\phi_{1}(\tau), ..., \phi_{M}(\tau)\bigr] \; = \;
     \int_{0}^{t} d\tau
   \Bigl[\frac{1}{2} \beta \sum_{a=1}^{M}\bigl(\partial_\tau \phi_{a}(\tau)\bigr)^2
   - \frac{1}{2} \beta^{2} u  \sum_{a,b=1}^{M} U\bigl(\phi_{a}(\tau) - \phi_{b}(\tau)\bigr) \Bigr];
\end{equation}
is the replica Hamiltonian with the attractive interaction potential $U(\phi)$ given in eq.(\ref{6}).
One can easily show that the function $\Psi\bigl(x_{1}, x_{2}, ..., x_{M}; \, t\bigr)$ 
is the wave function of one-dimensional quantum bosons which satisfy the imaginary time Schr\"odinger equation
\begin{equation}
 \label{40}
\beta \frac{\partial}{\partial t} \Psi({\bf x}; \, t) \; = \;
\frac{1}{2}\sum_{a=1}^{M} \, \frac{\partial^{2}}{\partial x_{a}^{2}} \Psi({\bf x}; \, t)
\; + \; \frac{1}{2} \, \beta^{3} u \, \sum_{a,b=1}^{M} U(x_{a} - x_{b}) \, \Psi({\bf x}; \, t)
\end{equation}
with the initial conditions $\Psi({\bf x}; \, 0) \; = \; \prod_{a=1}^{M}\, \delta(x_{a})$
(here we have introduced the vector notation ${\bf x} \equiv \{x_{1}, x_{2}, ..., x_{M}\}$).

The high temperature limit of the replica problem formulated above  is well
studied (for a review see e.g. \cite{rev} and references therein). It can be shown that in the limit
$\beta \to 0$ the interaction potential $U(x)$, eq.(\ref{6}), can be approximated by
the $\delta$-function, and in this case the generic solution of the Schr\"odinger equation
can be represented in terms of the Bethe ansatz eigenfunctions \cite{Lieb-Liniger,McGuire,Yang}.
However, at low temperatures, $T \lesssim \bigl(u R\bigr)^{1/3}$, the typical distance between
particles (defined by the wave function $\Psi({\bf x})$) becomes comparable with the size $R$
of the interaction potential $U(x)$ and its approximation by the $\delta$-function is no longer
valid. The zero temperature limit of the considered system has been studied in \cite{zero-T,Lecomte}

In the limit of low temperatures it is convenient to redefine the parameters of the system in the following way:
\begin{eqnarray}
\nonumber
\phi &=& R \, \tilde{\phi}
\\
 \label{41}
\beta &=& T_{*}^{-1} \tilde{\beta}
\\
\nonumber
\tau &=& \tau_{*} \tilde{\tau}
\end{eqnarray}
where
\begin{eqnarray}
 \label{42}
T_{*} &=& \Bigl(\frac{uR}{\sqrt{2\pi}}\Bigr)^{1/3}
\\
\nonumber
\\
\label{43}
\tau_{*} &=& \bigl(\sqrt{2\pi} R^{5} u^{-1} \bigr)^{1/3}
\end{eqnarray}
In the new notations the replica Hamiltonian (\ref{39}) reads
\begin{equation}
 \label{44}
\beta H_{M} \bigl[\tilde{\boldsymbol{\phi}} \bigr] \; = \;
     \int_{0}^{t/\tau_{*}} d\tilde{\tau}
   \Bigl[\; \frac{1}{2} \tilde{\beta} \sum_{a=1}^{M}\bigl(\partial_{\tilde{\tau}} \tilde{\phi}_{a}(\tilde{\tau})\bigr)^2
   - \frac{1}{2} \tilde{\beta}^{2} \sum_{a,b=1}^{M} U_{0}\bigl(\tilde{\phi}_{a}(\tilde{\tau}) - \tilde{\phi}_{b}(\tilde{\tau})\bigr) \; \Bigr];
\end{equation}
where
\begin{equation}
 \label{45}
U_{0}(\phi) \; = \; \exp\Bigl\{-\frac{1}{2} \phi^{2} \Bigr\}
\end{equation}
Accordingly, instead of eq.(\ref{40}) we get
\begin{equation}
 \label{46}
\tilde{\beta} \frac{\partial}{\partial \tilde{t}} \Psi({\bf \tilde{x}}; \, t) \; = \;
\frac{1}{2}\sum_{a=1}^{M} \, \frac{\partial^{2}}{\partial \tilde{x}_{a}^{2}} \Psi({\bf \tilde{x}}; \, t)
\; + \; \frac{1}{2} \, \tilde{\beta}^{3} \, \sum_{a,b=1}^{M} U_{0}(\tilde{x}_{a} - \tilde{x}_{b}) \, \Psi({\bf \tilde{x}}; \, t)
\end{equation}
where $\tilde{t} = t/\tau_{*}$ and $\tilde{x} = x/R$. Substituting here
$\Psi({\bf \tilde{x}}; \, t) \; = \; \psi({\bf \tilde{x}}) \, \exp\bigl\{-E \tilde{t}\bigr\}$
we obtain the following equation for the eigenfunctions $\psi({\bf \tilde{x}})$ and the eigenvalues (energy) $E$:
\begin{equation}
 \label{47}
-2\tilde{\beta} E \; \psi({\bf \tilde{x}}) \; = \;
\sum_{a=1}^{M} \, \frac{\partial^{2}}{\partial \tilde{x}_{a}^{2}} \psi({\bf \tilde{x}})
\; + \;  \tilde{\beta}^{3} \, \sum_{a,b=1}^{M} U_{0}(\tilde{x}_{a} - \tilde{x}_{b}) \, \psi({\bf \tilde{x}})
\end{equation}
which is controlled by the only parameter
\begin{equation}
 \label{48}
\tilde{\beta} \; = \; \beta \, T_{*} \; = \; \beta \, \bigl(uR\bigr)^{1/3} \, (2\pi)^{-1/6}
\end{equation}
We see that $T_{*}$, eq.(\ref{42}), is the crossover temperature which separates the high-temperatures, $T \gg T_{*}$,
and the low-temperatures, $T \ll T_{*}$, regimes. Note also that introduced above dimensionless inverse temperature
parameter $\tilde{\beta}$, eq.(\ref{48}), coincides with the Reynolds number $Re$, eq.(\ref{9}), so that the limit of
large Reynolds number in the Burgers problem corresponds to the zero temperature limit in the considered directed
polymers model.

Recently it has been demonstrated \cite{zero-T} that in the limit $\tilde{\beta}\to\infty$ the eigenfunction
$\psi({\bf \tilde{x}})$ acquires specific vector replica symmetry breaking (RSB) coordinate structure,
namely,  $M$ its arguments $\{\tilde{x}_{1}, ... ,\tilde{x}_{M}\}$  split into $K=M/m$ groups 
each consisting of $m$ particles.
In other words, to describe the coordinate structure of the eigenfunction $\psi({\bf \tilde{x}})$, instead of
the particles coordinates $\{\tilde{x}_{a}\} \; \; (a = 1, ..., M)$ one introduces the coordinates
of the center of masses of the groups $\bigl\{ X_{\alpha}\bigr\} \; \; (\alpha = 1, ..., K) $ and the
deviations $\{\xi^{\alpha}_{i}\} \; \; (i = 1, ..., m)$ of the particles of a given group $\alpha$ from
the position of its center of mass:
\begin{equation}
 \label{49}
\tilde{x}_{a} \; \to \; X_{\alpha} + \xi^{\alpha}_{i} \, ; \; \; \; \; \; \alpha = 1, ..., M/m \, ; \; \; \; i = 1, ..., m
\end{equation}
where $\sum_{i=1}^{m} \xi^{\alpha}_{i} = 0$. It can be shown \cite{zero-T} that in the zero temperature limit
the typical value of the deviations insides groups are small,
$\langle(\xi^{\alpha}_{i})^{2}\rangle\big|_{\tilde{\beta}\to\infty} \; \to \; 0$, while the typical distance
between the groups remains finite. As these two spatial scales are well separated,
the wave function $\psi({\bf \tilde{x}})$ factorizes into the product of two
contributions: the "external" wave function which depends only on the coordinates
$\{X_{\alpha}\}$ of the center of masses of the groups, and the "internal" wave functions which depend only on the
coordinates $\{\xi^{\alpha}_{i}\}$ of the particles inside the groups:
\begin{equation}
 \label{50}
\psi({\bf \tilde{x}}) \; \to \; \psi\bigl(X_{\alpha}; \; \xi^{\alpha}_{i}\bigr) \; \simeq \;
   \psi_{*}\bigl(X_{1}, ..., X_{M/m}\bigr) \times \prod_{\alpha=1}^{M/m} \psi_{0}\bigl(\xi^{\alpha}_{1}, ... \xi^{\alpha}_{m}\bigr)
\end{equation}
As the values $\xi^{\alpha}_{i}$ are small the interaction potential, eq.(\ref{45}), between the particles inside groups
can be approximated as
\begin{equation}
 \label{51}
U_{0}\bigl(\xi^{\alpha}_{i} - \xi^{\alpha}_{j}\bigr) \; \simeq \; 1 \; - \; \frac{1}{2} \bigl(\xi^{\alpha}_{i} - \xi^{\alpha}_{j}\bigr)^{2}
\end{equation}
Thus, according to eq.(\ref{47}), the corresponding equation for the "internal" eigenfunction
$\psi_{0}\bigl(\boldsymbol{\xi}\bigr)$ of any group reads
\begin{equation}
 \label{52}
-2\tilde{\beta} E_{0} \; \psi_{0}\bigl(\boldsymbol{\xi}\bigr) \; = \;
\sum_{i=1}^{m} \, \frac{\partial^{2}}{\partial \xi_{i}^{2}} \psi_{0}\bigl(\boldsymbol{\xi}\bigr)
\; + \; \tilde{\beta}^{3} m^{2} \psi_{0}\bigl(\boldsymbol{\xi}\bigr)
\; - \; \frac{1}{2} \tilde{\beta}^{3} \, \sum_{i,j=1}^{m} \bigl(\xi_{i} - \xi_{j}\bigr)^{2} \,
\psi_{0}\bigl(\boldsymbol{\xi}\bigr)
\end{equation}
where $\boldsymbol{\xi} \; = \; \{\xi_{1}, \xi_{2}, ..., \xi_{m}\}$.
One can easily show that this equation has the following exact (ground state) solution
\begin{equation}
 \label{53}
\psi_{0}\bigl(\boldsymbol{\xi}\bigr) \; = \;
     C \, \exp\Bigl\{
    -\frac{1}{4} \tilde{\beta}^{2} \bigl(\tilde{\beta} m\bigr)^{-1/2} \sum_{i,j=1}^{m} \bigl(\xi_{i} - \xi_{j}\bigr)^{2}
              \Bigr\}
\end{equation}
where $C$ is the normalization constant and
\begin{equation}
 \label{54}
E_{0} \; = \; -\frac{1}{2} \bigl(\tilde{\beta} m\bigr)^{2} \; + \; \frac{1}{2} (m-1) \sqrt{\tilde{\beta} m}
\end{equation}
is the ground state energy.

On the other hand, the "external" wave function $\psi_{*}\bigl({\bf X}\bigr)$
(with ${\bf X} = \{X_{1}, ..., X_{M/m} \}$) is defined by the equation
\begin{equation}
 \label{55}
-2 \bigl(\tilde{\beta} m\bigr) \, E_{*} \psi_{*}\bigl({\bf X}\bigr) \; = \;
\sum_{\alpha=1}^{M/m} \, \frac{\partial^{2}}{\partial X_{\alpha}^{2}} \psi_{*}\bigl({\bf X}\bigr) \; + \;
\frac{1}{2} \bigl(\tilde{\beta} m\bigr)^{3} \sum_{\alpha\not= \alpha'}^{M/m} U_{0}\bigl(X_{\alpha}-X_{\alpha'}\bigr)
\psi_{*}\bigl({\bf X}\bigr)
\end{equation}
In terms of the replica approach, the parameter $m$ of the RSB ansatz described above is an integer
such that $1 \leq m \leq M$  (so that $M/m$ is also an integer). In the framework of the standard
replica technique, after computing the corresponding partition function and its analytic continuation for
arbitrary (non-integer) values of $M$ and $m$, in the limit $M \to 0$ the parameter $m$ takes
continuous (real) values at the interval $0 \leq m \leq 1$. Its actual physical value $m(\tilde{\beta})$
is fixed by the condition of the {\it maximum} of the total (linear in time $t \to \infty$) replica
free energy. It can be shown \cite{zero-T} that in the limit $\tilde{\beta} \to \infty$ the value $m(\tilde{\beta})$
is defined by the relation
\begin{equation}
 \label{56}
\tilde{\beta} \, m \; = \; \zeta_{0}
\end{equation}
where $\zeta_{0}$ is a number of the order of one (such that $m(\tilde{\beta}) \to 0$ as
$\tilde{\beta} \to \infty$). The exact value of $\zeta_{0}$ is yet to be computed, as it is
defined by the exact solution of the "external" problem, eq.(\ref{55}), which at present is not known.

In terms of this RSB ansatz in the zero temperature limit the replica partition function of the considered system,
eq.(\ref{31}), factorizes into two parts:
\begin{equation}
 \label{57}
Z_{x,\epsilon,t} \bigl(M,N_{1},N_{2},N_{3} \bigr) \; \simeq \;
Z_{*}\bigl[(\tilde{\beta} m), \; M/m, \; \tilde{t}\bigr] \times
Z_{0}\bigl(M, m, \tilde{\beta},  N_{1}, N_{2}, N_{3}, x, \epsilon\bigr)
\end{equation}
where $Z_{*}$ is the "external" replica partition function:
\begin{equation}
 \label{58}
Z_{*}  =  \prod_{\alpha=1}^{M/m}
      \Biggl[\int_{\varphi_{\alpha}(0)=0}^{\varphi_{\alpha}(\tilde{t})=0} {\cal D}\varphi_{\alpha}(\tau)\Biggr]
         \exp\Biggl\{-\frac{1}{2} \int_{0}^{\tilde{t}} d\tau
    \Bigl[(\tilde{\beta} m) \sum_{\alpha=1}^{M/m} \bigl(\partial_{\tau} \varphi_{\alpha}\bigr)^{2}
           - (\tilde{\beta} m)^{2} \sum_{\alpha\not= \alpha'}^{M/m}
               U_{0}\bigl(\varphi_{\alpha} - \varphi_{\alpha'}\bigr) \Bigr] -
                 \tilde{t} \frac{M}{m} \, E_{0} \Biggr\}
\end{equation}
where $E_{0}$ is given in eq.(\ref{54}). Note that in the limit $\tilde{t}\to\infty$ this partition function
is getting independent of $x$ and $\epsilon$, as these parameters are not scaling with $\tilde{t}$.
The above "external" partition function $Z_{*}$ defines the extensive in $\tilde{t}\to\infty$ part of the directed polymer
free energy and fixes the value of the parameter $m = m(\tilde{\beta})$, eq.(\ref{56}), and it is in this way that
the parameters of the large-scale random potential influence the small-scale statistics defined by the "internal"
partition function (see below) which also depends on the value of $m(\tilde{\beta})$.
 On the other hand, by definition,
\begin{equation}
 \label{59}
\lim_{M\to 0} \, Z_{*}\bigl[(\tilde{\beta} m), \; M/m, \; \tilde{t}\bigr] \; = \; 1
\end{equation}
and therefore, except for fixing the value of the replica parameter $m(\tilde{\beta})$,
this part of the total partition function does not contribute to the probability
function  $ {\cal W}_{x,\epsilon}\bigl(f_{1}, f_{2},f_{3}\bigr)$, eqs.(\ref{33}) and (\ref{30}).
This probability function is defined only by the "internal" (independent of $\tilde{t}$) partition function
\begin{eqnarray}
\nonumber
{\cal Z}_{0}\bigl(N_{1}, N_{2}, N_{3}; x, \epsilon\bigr) &=&
\lim_{\tilde{\beta}\to\infty} \lim_{M\to 0} Z_{0}\bigl(M, m, \tilde{\beta},  N_{1}, N_{2}, N_{3}, x, \epsilon\bigr)
\\
\nonumber
\\
&=& \lim_{\tilde{\beta}\to\infty} \lim_{M\to 0} \Biggl[ \sum_{\{\tilde{\xi}^{\alpha}_{i}\}}
   \prod_{\alpha=1}^{M/m}
\psi_{0}\bigl(\tilde{\xi}^{\alpha}_{1}, ..., \tilde{\xi}^{\alpha}_{m}\bigr)\Big|_{\{\tilde{\xi}^{\alpha}_{i}\} =
(\tilde{x}/2; \; \tilde{x}/2 - \epsilon ; \; -\tilde{x}/2 + \epsilon ; \; -\tilde{x}/2)} \Biggr]
 \label{60}
\end{eqnarray}
where the explicit expression for $\psi_{0}\bigl(\boldsymbol{\tilde{\xi}}\bigr)$ is given in eq.(\ref{53}), and where
we have to sum over all possible distributions of
$M$ particle coordinates $\{\tilde{\xi}^{\alpha}_{i}\} \; \; (\alpha = 1, ..., M/m ; \; \; i = 1, ..., m)$
over four end-points $\tilde{x}/2; \; \tilde{x}/2 - \epsilon ; \; -\tilde{x}/2 + \epsilon$ and $ -\tilde{x}/2$
with $\tilde{x} = x/R$ and $\tilde{\xi}^{\alpha}_{i}  = \xi^{\alpha}_{i}/R$.

\section{Free energies probability distribution function}

Substituting eqs.(\ref{53}) and  (\ref{48})  as well as $\tilde{x} = x/R$ and $\tilde{\xi}^{\alpha}_{i}  = \xi^{\alpha}_{i}/R$
into eq.(\ref{60}) we gets
\begin{equation}
 \label{61}
{\cal Z}_{0}\bigl(N_{1}, N_{2}, N_{3}; x, \epsilon\bigr) \; = \;
     \lim_{\beta\to\infty} \lim_{M\to 0} \Biggl[ \sum_{\{\tilde{\xi}^{\alpha}_{i}\}}
   \prod_{\alpha=1}^{M/m}
    \exp\Bigl\{ -\frac{1}{4} \beta^{2} \gamma^{2} \sum_{i,j=1}^{m} \bigl(\xi^{\alpha}_{i} - \xi^{\alpha}_{j}\bigr)^{2}
              \Bigr\}\Big|_{\{\xi^{\alpha}_{i}\} =
(x/2; \; x/2 - \epsilon ; \; -x/2 + \epsilon ; \; -x/2)} \Biggr]
\end{equation}
where
\begin{equation}
 \label{62}
\gamma \; = \; \frac{T_{*}}{R (\tilde{\beta} m)^{1/4}}
\end{equation}
and $T_{*}$ is given in eq.(\ref{42}). Note that the normalization factor $C$ of the wave function
(\ref{53}) can be dropped out in eq.(\ref{61}), as $\lim_{M\to 0} C^{M/m} = 1$.
According to the definition, eq.(\ref{31}), in the summation over
various distributions of $M$ end-points $\xi^{\alpha}_{i}$ over four spatial points the total number
of $\xi^{\alpha}_{i}$'s attached to
$x/2$, $ x/2 - \epsilon$, $-x/2 + \epsilon$ and $-x/2$ are equal to $N_{1}$, $N_{2}$, $N_{3}$ and
$(M-N_{1}-N_{2}-N_{3})$ correspondingly. Let us denote the number of $\xi^{\alpha}_{i}$'s of the group $\alpha$
attached to the points $x/2$, $ x/2 - \epsilon$, $-x/2 + \epsilon$ and $-x/2$ by
$k^{\alpha}_{1}$, $k^{\alpha}_{2}$, $k^{\alpha}_{3}$ and $k^{\alpha}_{4}$. As the total number of particles
in each group is equal to $m$, by definition,
\begin{equation}
 \label{63}
k^{\alpha}_{1} + k^{\alpha}_{2} + k^{\alpha}_{3} + k^{\alpha}_{4} \; = \; m
\end{equation}
and
\begin{equation}
 \label{64}
\left\{
     \begin{array}{ll}
          \sum_{\alpha=1}^{M/m} k^{\alpha}_{1} \; = \; N_{1}
       \\
       \\
           \sum_{\alpha=1}^{M/m} k^{\alpha}_{2} \; = \; N_{2}
       \\
       \\
           \sum_{\alpha=1}^{M/m} k^{\alpha}_{3} \; = \; N_{3}
       \\
       \\
           \sum_{\alpha=1}^{M/m} k^{\alpha}_{4} \; = \; M-N_{1}-N_{2}-N_{3}
    \end{array}
\right.
\end{equation}
\begin{figure}[h]
\begin{center}
   \includegraphics[width=13.0cm]{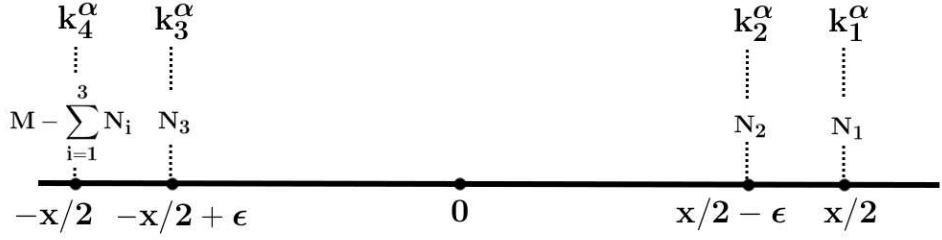}
\caption[]{Schematic representation of the replica structure of the partition function in eqs.(\ref{61})-(\ref{65}).}
\end{center}
\label{figure2}
\end{figure}

Schematically the above replica structure of the partition function (\ref{61}) is represented in Fig.2.
Accordingly, the factors $\bigl(\xi^{\alpha}_{i} - \xi^{\alpha}_{j}\bigr)^{2}$ in eq.(\ref{61})
can take four possible values: $\epsilon^{2}$, $(x-\epsilon)^{2}$, $(x-2\epsilon)^{2}$ and $x^{2}$.
Simple combinatoric considerations yield:
\begin{eqnarray}
 \nonumber
{\cal Z}_{0}\bigl(N_{1}, N_{2}, N_{3}; x, \epsilon\bigr) &=&
     \lim_{\beta\to\infty} \lim_{M\to 0}
\Biggl\{
\frac{N_{1}! \, N_{2}! \, N_{3}! \, (M - N_{1}- N_{2}-N_{3})!}{M!}
\times
\\
\nonumber
\\
\label{65}
&\times&
\prod_{\alpha=1}^{M/m}
\Biggl[
\Bigl(\prod_{i=1}^{4} \sum_{k^{\alpha}_{i}=0}^{m} \Bigr)
\frac{m!}{k^{\alpha}_{1}! \, k^{\alpha}_{2}! \, k^{\alpha}_{3}! \, k^{\alpha}_{4}!}
\boldsymbol{\delta}\Bigl(\sum_{i=1}^{4}k^{\alpha}_{i}, \; m\Bigr)
\exp\Bigl\{-\frac{1}{4} \beta^{2} \gamma^{2} \sum_{i,j=1}^{4} D_{ij} k^{\alpha}_{i}k^{\alpha}_{j} \Bigr\}
\Biggr]
\times
\\
\nonumber
\\
\nonumber
&\times&
\boldsymbol{\delta}\Bigl(\sum_{\alpha=1}^{M/m}k^{\alpha}_{1}, \; N_{1}\Bigr) \;
\boldsymbol{\delta}\Bigl(\sum_{\alpha=1}^{M/m}k^{\alpha}_{2}, \; N_{2}\Bigr) \;
\boldsymbol{\delta}\Bigl(\sum_{\alpha=1}^{M/m}k^{\alpha}_{3}, \; N_{3}\Bigr)
\Biggr\}
\end{eqnarray}
where $\boldsymbol{\delta}(p,q)$ is the Kronecker symbol and
\begin{equation}
 \label{66}
\hat{D}  \; = \; \left( \begin{array}{cccc}
                   0                & \epsilon^{2}      & (x-\epsilon)^{2}  & x^{2}            \\
                   \epsilon^{2}     &  0                & (x-2\epsilon)^{2} & (x-\epsilon)^{2}\\
                   (x-\epsilon)^{2} & (x-2\epsilon)^{2} & 0                 & \epsilon^{2}     \\
                   x^{2}            & (x-\epsilon)^{2}  & \epsilon^{2}      & 0                \\
                  \end{array}
                 \right)
\end{equation}

Note that the last constraint in eq.(\ref{64})
can be dropped out of the expression (\ref{65}), as it is automatically fulfilled due to the previous three ones
together with the condition (\ref{63}).

Substituting the matrix (\ref{66}) into eq.(\ref{65}) we get
\begin{eqnarray}
 \nonumber
{\cal Z}_{0}\bigl(N_{1}, N_{2}, N_{3}; x, \epsilon\bigr) &=&
     \lim_{\beta\to\infty} \lim_{M\to 0}
\Biggl\{
\frac{N_{1}! \, N_{2}! \, N_{3}! \, (M - N_{1}- N_{2}-N_{3})!}{M!}
\times
\\
\nonumber
\\
\label{67}
&\times&
\exp\Bigl\{
-\frac{1}{2} \beta N_{1} (\beta m) \gamma^{2} x^{2}
-\frac{1}{2} \beta N_{2} (\beta m) \gamma^{2} (x-\epsilon)^{2}
-\frac{1}{2} \beta N_{3} (\beta m) \gamma^{2} \epsilon^{2}
\Bigr\}
\times first
\\
\nonumber
\\
\nonumber
&\times&
\prod_{\alpha=1}^{M/m}
\Biggl[
\sum_{k^{\alpha}_{1},k^{\alpha}_{2},k^{\alpha}_{3}=0}^{m} \; C^{m}_{k^{\alpha}_{1},k^{\alpha}_{2},k^{\alpha}_{3}}
\exp\Bigl\{
\frac{1}{2} \beta^{2} \gamma^{2} \bigl(x k^{\alpha}_{1} + (x-\epsilon) k^{\alpha}_{2} + \epsilon k^{\alpha}_{3}\bigr)^{2}
\Bigr\}
\Biggr]
\times
\\
\nonumber
\\
\nonumber
&\times&
\boldsymbol{\delta}\Bigl(\sum_{\alpha=1}^{M/m}k^{\alpha}_{1}, \; N_{1}\Bigr) \;
\boldsymbol{\delta}\Bigl(\sum_{\alpha=1}^{M/m}k^{\alpha}_{2}, \; N_{2}\Bigr) \;
\boldsymbol{\delta}\Bigl(\sum_{\alpha=1}^{M/m}k^{\alpha}_{3}, \; N_{3}\Bigr)
\Biggr\}
\end{eqnarray}
where
\begin{equation}
 \label{68}
C^{m}_{k^{\alpha}_{1},k^{\alpha}_{2},k^{\alpha}_{3}} \; = \;
    \frac{m!}{k^{\alpha}_{1}! \, k^{\alpha}_{2}! \, k^{\alpha}_{3}! \, \bigl(m-k^{\alpha}_{1}-k^{\alpha}_{2}-k^{\alpha}_{3}\bigr)!}
\end{equation}
Using the standard integral representation of the Kronecker symbol,
\begin{equation}
 \label{69}
 \boldsymbol{\delta}(p, \, q) \; = \; \oint \frac{dz}{2\pi i z} \, z^{p-q}
\end{equation}
(where contour of integration in the complex plane is the circle around zero) the partition
function, eq.(\ref{67}), can be represented as follows:
\begin{eqnarray}
 \nonumber
{\cal Z}_{0}\bigl(N_{1}, N_{2}, N_{3}; x, \epsilon\bigr) &=&
     \lim_{\beta\to\infty} \lim_{M\to 0}
\Biggl\{
\frac{N_{1}! \, N_{2}! \, N_{3}! \, (M - N_{1}- N_{2}-N_{3})!}{M!}
\exp\bigl\{
-\beta N_{1} f_{01} -\beta N_{2} f_{02}-\beta N_{1} f_{03}
\bigr\}
\times
\\
\nonumber
\\
\nonumber
&\times&
\frac{1}{(2\pi i)^{3}} \oint \frac{dz_{1}}{z_{1}} \, z_{1}^{-N_{1}}
                       \oint \frac{dz_{2}}{z_{2}} \, z_{2}^{-N_{2}}
                       \oint \frac{dz_{3}}{z_{3}} \, z_{3}^{-N_{3}}
\times
\\
\nonumber
\\
\label{70}
&\times&
\Biggl[
\Biggl<\Bigl(
1 + z_{1}\exp\{\beta\gamma x \xi\}  + z_{2}\exp\{\beta\gamma (x-\epsilon) \xi\} + z_{1}\exp\{\beta\gamma \epsilon \xi\}
\Bigr)^{m}\Biggr>_{\xi}
\Biggr]^{M/m}\; \;
\Biggr\}
\end{eqnarray}
where
\begin{eqnarray}
 \nonumber
f_{01} &=& \frac{1}{2} (\beta m) \gamma^{2} x^{2}
\\
\label{71}
f_{01} &=& \frac{1}{2} (\beta m) \gamma^{2} (x-\epsilon)^{2}
\\
\nonumber
f_{03} &=& \frac{1}{2} (\beta m) \gamma^{2} \epsilon^{2}
\end{eqnarray}
and $\bigl< (...)\bigr>_{\xi}$ denotes the Gaussian average over the variable $\xi$:
\begin{equation}
 \label{72}
   \bigl<(...)\bigr>_{\xi} \; \equiv \; \int_{-\infty}^{+\infty}
              \frac{d\xi}{\sqrt{2\pi}} \; (...) \exp\Bigl\{-\frac{1}{2} \xi^{2} \Bigr\}
\end{equation}
Now the expression for the replica partition function, eq.(\ref{70}), can be analytically continued
for arbitrary non-integer values of the parameter $M$. In particular, the factorial prefactor
\begin{equation}
 \label{73}
\frac{(M - N_{1}- N_{2}-N_{3})!}{M!} \; \to \; \frac{\Gamma(M - N_{1}- N_{2}-N_{3} + 1)}{\Gamma(M+1)}
\end{equation}
Using the Gamma function relation,
\begin{equation}
\label{74}
\Gamma(-z) \; = \; -\frac{\pi}{\Gamma(z+1) \sin(\pi z)}
\end{equation}
for {\it integer and  positive} values of $N_{1,2,3}$ this prefactor can be represented as follows,
\begin{eqnarray}
 \nonumber
\frac{\Gamma(M - N_{1}- N_{2}-N_{3} + 1)}{\Gamma(M+1)} &=&
-\frac{\pi}{\Gamma(M+1) \Gamma(N_{1}+N_{2}+N_{3}-M) \sin\bigl[\pi(N_{1}+N_{2}+N_{3}-1)-\pi M\bigr]}
\\
\nonumber
\\
\label{75}
&=& \frac{\pi \, (-1)^{N_{1}+N_{2}+N_{3}-1}}{\sin\bigl(\pi M\bigr) \,  \Gamma(N_{1}+N_{2}+N_{3}-M) \, \Gamma(M+1)}
\end{eqnarray}
so that in the limit $M \to 0$ we get
\begin{equation}
 \label{76}
\frac{\pi \, (-1)^{N_{1}+N_{2}+N_{3}-1}}{\sin\bigl(\pi M\bigr) \,  \Gamma(N_{1}+N_{2}+N_{3}-M) \, \Gamma(M+1)}\Bigg|_{M\to 0}
\; \to \;
\frac{(-1)^{N_{1}+N_{2}+N_{3}-1}}{M \;  \Gamma(N_{1}+N_{2}+N_{3})}
\end{equation}
On the other hand, for the last factor in the expression (\ref{70}) we find
\begin{equation}
 \label{77}
\bigl[...\bigr]^{M/m}\Big|_{M \to 0} \; \to \; 1 \; + \; \frac{M}{m} \ln\bigl[...\bigr]
\end{equation}
Substituting eqs(\ref{76}) and (\ref{77}) into eq.(\ref{70}) and
taking into account that for any nonzero {\it integer} $N$,
\begin{equation}
 \label{78}
 \oint \frac{dz_{1}}{z} \, z^{-N} \; = \; 0
\end{equation}
in the limit $M \to 0$ we get
\begin{eqnarray}
 \nonumber
{\cal Z}_{0}\bigl(N_{1}, N_{2}, N_{3}; x, \epsilon\bigr) &=&
     \lim_{\beta\to\infty}
\Biggl\{
\frac{(-1)^{N_{1}+N_{2}+N_{3}-1} \Gamma(N_{1}+1) \, \Gamma(N_{2}+1) \, \Gamma(N_{3}+1)}{
m \, \Gamma(N_{1}+N_{2}+N_{3})}
\exp\bigl\{
-\beta N_{1} f_{01} -\beta N_{2} f_{02}-\beta N_{1} f_{03}
\bigr\}
\times
\\
\nonumber
\\
\nonumber
&\times&
\frac{1}{(2\pi i)^{3}} \oint \frac{dz_{1}}{z_{1}} \, z_{1}^{-N_{1}}
                       \oint \frac{dz_{2}}{z_{2}} \, z_{2}^{-N_{2}}
                       \oint \frac{dz_{3}}{z_{3}} \, z_{3}^{-N_{3}}
\times
\\
\nonumber
\\
\label{79}
&\times&
\ln \Biggl[
\Biggl<\Bigl(
1 + z_{1}\exp\{\beta\gamma x \xi\}  + z_{2}\exp\{\beta\gamma (x-\epsilon) \xi\} + z_{1}\exp\{\beta\gamma \epsilon \xi\}
\Bigr)^{m}\Biggr>_{\xi}
\Biggr]\; \;
\Biggr\}
\end{eqnarray}
Substituting this expression into eqs.(\ref{30}) and (\ref{33}) for the free energy probability distribution function
we obtain
\begin{eqnarray}
 \nonumber
{\cal W}_{x,\epsilon}\bigl(f_{1}, f_{2},f_{3}\bigr) &=& \lim_{\beta\to\infty} \Biggl\{
\sum_{N_{1},N_{2},N_{3}=1}^{\infty}
\frac{\exp\bigl\{
\beta N_{1}(f_{1}- f_{01}) +\beta N_{2}(f_{1}- f_{02}) + \beta N_{1}(f_{1}- f_{03})
\bigr\}}{
m \, \Gamma(N_{1}+N_{2}+N_{3})}
\times
\\
\nonumber
\\
\nonumber
&\times&
\frac{1}{(2\pi i)^{3}} \oint \frac{dz_{1}}{z_{1}} \, z_{1}^{-N_{1}}
                       \oint \frac{dz_{2}}{z_{2}} \, z_{2}^{-N_{2}}
                       \oint \frac{dz_{3}}{z_{3}} \, z_{3}^{-N_{3}}
\times
\\
\nonumber
\\
\label{80}
&\times&
\ln \Biggl[
\Biggl<\Bigl(
1 + z_{1}\exp\{\beta\gamma x \xi\}  + z_{2}\exp\{\beta\gamma (x-\epsilon) \xi\} + z_{1}\exp\{\beta\gamma \epsilon \xi\}
\Bigr)^{m}\Biggr>_{\xi}
\Biggr]\; \;
\Biggr\}
\end{eqnarray}
The limit $\beta\to \infty$ is somewhat tricky: on one hand, according to eq.(\ref{56}) in the zero temperature limit $m \propto 1/\beta \to 0$ and on the other hand we have several  exponential factors in
the above expression which are formally divergent in this limit.
To take the limit $m \propto 1/\beta \to 0$ the expression under the logarithm in eq.(\ref{80})
can be represented as follows:
\begin{eqnarray}
\nonumber
&&\Biggl<\Bigl(
1 + z_{1}\exp\{\beta\gamma x \xi\}  + z_{2}\exp\{\beta\gamma (x-\epsilon) \xi\} + z_{1}\exp\{\beta\gamma \epsilon \xi\}
\Bigr)^{m}\Biggr>_{\xi} \; = \;
\\
\nonumber
\\
\label{81}
&=&
1 + \sum_{k_{1}+k_{2}+k_{3}\geq 1}^{\infty} C^{m}_{k_{1}k_{2}k_{3}} \;
   z_{1}^{k_{1}}z_{2}^{k_{2}}z_{3}^{k_{3}} \;
\Bigl<
\exp\bigl\{
\beta k_{1}\gamma x \xi + \beta k_{2}\gamma (x-\epsilon) \xi + \beta k_{3}\gamma\epsilon \xi
\bigr\}
\Bigr>_{\xi}
\end{eqnarray}
where
\begin{equation}
\label{82}
C^{m}_{k_{1}k_{2}k_{3}} \; = \; \frac{\Gamma(m+1)}{
\Gamma(k_{1}+1) \Gamma(k_{2}+1) \Gamma(k_{3}+1) \Gamma(m - k_{1}- k_{2}- k_{3} +1)}
\end{equation}
In the limit $m\to 0$ we get (sf eqs.(\ref{75})-(\ref{76}))
\begin{equation}
\label{83}
C^{m}_{k_{1}k_{2}k_{3}}\Big|_{m\to 0} \simeq
m \frac{(-1)^{k_{1}+ k_{2}+ k_{3} -1}}{k_{1}+ k_{2}+ k_{3}} \;
C^{0}_{k_{1}k_{2}k_{3}}
\end{equation}
where
\begin{equation}
\label{84}
C^{0}_{k_{1}k_{2}k_{3}} \; = \;
\frac{\Gamma(k_{1}+ k_{2}+ k_{3}+1)}{\Gamma(k_{1}+1) \Gamma(k_{2}+1) \Gamma(k_{3}+1)}
\end{equation}
Substituting eqs.(\ref{83}) and (\ref{81}) into eq.(\ref{80}) and expending the logarithm term
after integrations over $z_{1}$, $z_{2}$ and $z_{3}$ we obtain
\begin{eqnarray}
 \nonumber
{\cal W}_{x,\epsilon}\bigl(f_{1}, f_{2},f_{3}\bigr) &=& \lim_{\beta\to\infty} \Biggl\{
\frac{1}{(\beta m)}
\sum_{n=1}^{\infty} \frac{(-1)^{n-1}}{n}
\times
\\
\nonumber
\\
\nonumber
&\times&
\prod_{\alpha=1}^{n} \Biggl[
(\beta m)
\sum_{k_{1}^{\alpha}+k_{2}^{\alpha}+k_{3}^{\alpha}\geq 1}^{\infty}
\frac{(-1)^{k_{1}^{\alpha}+ k_{2}^{\alpha}+ k_{3}^{\alpha} -1}}{
\beta(k_{1}^{\alpha}+ k_{2}^{\alpha}+ k_{3}^{\alpha})} \;
C^{0}_{k^{\alpha}_{1}k^{\alpha}_{2}k^{\alpha}_{3}}
\Bigl<
\exp\bigl\{
\beta k_{1}^{\alpha}\gamma x \xi + \beta k_{2}^{\alpha}\gamma (x-\epsilon) \xi + \beta k_{3}^{\alpha}\gamma\epsilon \xi
\bigr\}
\Bigr>_{\xi}
\Biggr]
\times
\\
\nonumber
\\
\nonumber
&\times&
\sum_{N_{1},N_{2},N_{3}=1}^{\infty}
\frac{\beta(N_{1}+N_{2}+N_{3})}{\Gamma(N_{1}+N_{2}+N_{3}+1)}
\exp\bigl\{
\beta N_{1}(f_{1}- f_{01}) +\beta N_{2}(f_{1}- f_{02}) + \beta N_{1}(f_{1}- f_{03})
\bigr\}
\times
\\
\nonumber
\\
\label{85}
&\times&
\boldsymbol{\delta}\Bigl(\sum_{\alpha=1}^{n}k^{\alpha}_{1}, \; N_{1}\Bigr) \;
\boldsymbol{\delta}\Bigl(\sum_{\alpha=1}^{n}k^{\alpha}_{2}, \; N_{2}\Bigr) \;
\boldsymbol{\delta}\Bigl(\sum_{\alpha=1}^{n}k^{\alpha}_{3}, \; N_{3}\Bigr)
\Biggr\}
\end{eqnarray}
Substituting here $\beta m \; = \; \tilde{\beta} m /T_{*} \; = \; \zeta_{0}/T_{*}$
(see eqs.(\ref{56}), (\ref{41}) and (\ref{42}))
and resolving the Kronecker symbols in the summations over $N_{1}$, $N_{2}$ and $N_{3}$
we get
\begin{eqnarray}
 \nonumber
&&{\cal W}_{x,\epsilon}\bigl(f_{1}, f_{2},f_{3}\bigr) \; = \; \frac{T_{*}}{\zeta_{0}} \lim_{\beta\to\infty} \Biggl\{
\sum_{n=1}^{\infty} \frac{(-1)^{n-1}}{n} \; \Bigl(\frac{\zeta_{0}}{T_{*}}\Bigr)^{n}
\prod_{\alpha=1}^{n} \Biggl[
\sum_{k_{1}^{\alpha}+k_{2}^{\alpha}+k_{3}^{\alpha}\geq 1}^{\infty}
\frac{(-1)^{k_{1}^{\alpha}+ k_{2}^{\alpha}+ k_{3}^{\alpha} -1}}{
\beta(k_{1}^{\alpha}+ k_{2}^{\alpha}+ k_{3}^{\alpha})} \;
C^{0}_{k^{\alpha}_{1}k^{\alpha}_{2}k^{\alpha}_{3}}
\times
\\
\nonumber
\\
\nonumber
&\times&
\Bigl<
\exp\Bigl\{
\beta k_{1}^{\alpha}\bigl(\gamma x \xi + f_{1} - f_{01}\bigr) +
\beta k_{2}^{\alpha}\bigl(\gamma (x-\epsilon) \xi + f_{2}-f_{02}\bigr) +
\beta k_{3}^{\alpha}\bigl(\gamma\epsilon \xi + f_{3} - f_{03}\bigr)
\Bigr\}
\Bigr>_{\xi}
\Biggr]
\times
\\
\nonumber
\\
\label{86}
&\times&
\frac{\beta\sum_{\alpha=1}^{n}\bigl(k_{1}^{\alpha}+k_{2}^{\alpha}+k_{3}^{\alpha}\bigr)}{
\Gamma\Bigl[\sum_{\alpha=1}^{n}\bigl(k_{1}^{\alpha}+k_{2}^{\alpha}+k_{3}^{\alpha}\bigr) + 1\Bigr]} \;
\boldsymbol{\theta}\Bigl(\sum_{\alpha=1}^{n}k^{\alpha}_{1}\, - \, 1\Bigr) \;
\boldsymbol{\theta}\Bigl(\sum_{\alpha=1}^{n}k^{\alpha}_{2}\, - \, 1\Bigr) \;
\boldsymbol{\theta}\Bigl(\sum_{\alpha=1}^{n}k^{\alpha}_{3}\, - \, 1\Bigr)
\Biggr\}
\end{eqnarray}
where the symbol $\boldsymbol{\theta}(p\, - \, 1)$ (the "discrete step function") indicates that $p \geq 1$.
Note however, that according to eq.(\ref{86})  the contributions with $\sum_{\alpha=1}^{n}k^{\alpha}_{i} = 0$
(such that all $k^{1}_{i} = k^{2}_{1} = ... = k^{n}_{i} = 0$) are independent of the corresponding
free energy parameter $f_{i}$. On the other hand, the probability density function $P_{x,\epsilon}(f_{1}, f_{2}, f_{3})$
which we are aiming to derive is given by the derivatives of the above probability function
$W$ over {\it all three} variables $f_{1}$, $f_{2}$ and $f_{3}$ (see eq.(\ref{32})). Therefore, as far as
the PDF $P_{x,\epsilon}(f_{1}, f_{2}, f_{3})$ is concerned the restrictions imposed by the
last three "discrete step function" in eq.(\ref{86}) can be omitted.

  The factor $\beta\sum_{\alpha=1}^{n}\bigl(k_{1}^{\alpha}+k_{2}^{\alpha}+k_{3}^{\alpha}\bigr)$
in the numerator of the last term in eq.(\ref{86}) can be obtained by taking the derivatives
$\Bigl(\frac{\partial}{\partial f_{1}} + \frac{\partial}{\partial f_{2}} +\frac{\partial}{\partial f_{3}}\Bigr)$
of ${\cal W}_{x,\epsilon}\bigl(f_{1}, f_{2},f_{3}\bigr)$. On the other hand,
\begin{equation}
 \label{87}
\frac{1}{\beta \bigl(k_{1}^{\alpha}+k_{2}^{\alpha}+k_{3}^{\alpha}\bigr)} \; = \;
\int_{0}^{+\infty} dy \exp\bigl\{ - \beta \bigl(k_{1}^{\alpha}+k_{2}^{\alpha}+k_{3}^{\alpha}\bigr) \; y \bigr\}
\end{equation}
Substituting this into eq.(\ref{86}) and then, substituting the obtained expression into eq.(\ref{32}) for the PDF
$P_{x,\epsilon}(f_{1}, f_{2}, f_{3})$ we get
\begin{eqnarray}
 \nonumber
&&P_{x,\epsilon}(f_{1}, f_{2}, f_{3}) \; = \;
\frac{T_{*}}{\zeta_{0}}
\frac{\partial^{3}}{\partial f_{1} \; \partial f_{2} \; \partial f_{3}}
\Bigl(\frac{\partial}{\partial f_{1}} + \frac{\partial}{\partial f_{2}} +\frac{\partial}{\partial f_{3}}\Bigr)
\lim_{\beta\to\infty} \Biggl\{
\sum_{n=1}^{\infty} \frac{(-1)^{n-1}}{n} \; \Bigl(\frac{\zeta_{0}}{T_{*}}\Bigr)^{n}
\times
\\
\nonumber
\\
\nonumber
&\times&
\prod_{\alpha=1}^{n} \Biggl[
\int_{0}^{+\infty} dy
\sum_{k_{1}^{\alpha}+k_{2}^{\alpha}+k_{3}^{\alpha}\geq 1}^{\infty}
(-1)^{k_{1}^{\alpha}+ k_{2}^{\alpha}+ k_{3}^{\alpha} -1}\;
C^{0}_{k^{\alpha}_{1}k^{\alpha}_{2}k^{\alpha}_{3}}
\times
\\
\nonumber
\\
\nonumber
&\times&
\Bigl<
\exp\Bigl\{
\beta k_{1}^{\alpha}\bigl(\gamma x \xi + f_{1} - f_{01}-y\bigr) +
\beta k_{2}^{\alpha}\bigl(\gamma (x-\epsilon) \xi + f_{2}-f_{02}-y\bigr) +
\beta k_{3}^{\alpha}\bigl(\gamma\epsilon \xi + f_{3} - f_{03}-y\bigr)
\Bigr\}
\Bigr>_{\xi}
\Biggr]
\times
\\
\nonumber
\\
\label{88}
&\times&
\frac{1}{\Gamma\Bigl[\sum_{\alpha=1}^{n}\bigl(k_{1}^{\alpha}+k_{2}^{\alpha}+k_{3}^{\alpha}\bigr) + 1\Bigr]} \;
\Biggr\}
\end{eqnarray}
In the limit $\beta\to\infty$ the summation the series over $k^{\alpha}_{i}$ in the above expression can be done
using the their integral representation. Namely, let us consider the series of a general type
\begin{equation}
 \label{89}
R(\beta) \; = \; \sum_{k=0}^{\infty} (-1)^{k-1} \; \Phi\bigl(\beta k; \; k\bigr)
\end{equation}
where $\Phi(z, z')$ is a "good" analytic function in the complex plane.
One can easily see that the summation  in eq.(\ref{89}) can be changed by the integration in the complex plane:
\begin{equation}
 \label{90}
R(\beta) \; = \; \frac{1}{2i} \int_{{\cal C}} \frac{dz}{\sin(\pi z)} \; \Phi\bigl(\beta z; \; z\bigr)
\end{equation}
where the integration goes over the contour ${\cal C}$ shown in Fig.3, and it is assumed that the function $\Phi$ is such that
its integration at infinity gives no contribution. Indeed, due to the sign alternating contributions of simple
poles at integer $z = 1, 2, ...$ eq.(\ref{90}) reduces to eq.(\ref{89}). Then, redefining $z \to z/\beta$ we gets
\begin{equation}
 \label{91}
\lim_{\beta\to\infty} R(\beta) \; = \; \frac{1}{2\pi i} \int_{{\cal C}} \frac{dz}{z} \;
       \lim_{\beta\to\infty} \Phi\bigl(z; \; z/\beta\bigr)
\end{equation}
\begin{figure}[h]
\begin{center}
   \includegraphics[width=8.0cm]{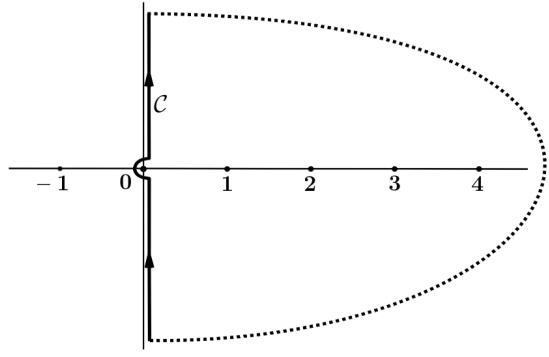}
\caption[]{The contour of integration in  eq.(\ref{90})}
\end{center}
\label{figure3}
\end{figure}
In terms of the above integral representation, changing $k^{\alpha}_{i} \to z^{\alpha}_{i}/\beta$
for the Gamma function factors in eq.(\ref{88})  we have:
\begin{equation}
  \label{92}
\Gamma\Bigl[\sum_{\alpha=1}^{n}\bigl(k_{1}^{\alpha}+k_{2}^{\alpha}+k_{3}^{\alpha}\bigr) + 1\Bigr] \; \to \;
\Gamma\Bigl[\sum_{\alpha=1}^{n}\bigl(z_{1}^{\alpha}+z_{2}^{\alpha}+z_{3}^{\alpha}\bigr)/\beta + 1\Bigr]\Big|_{\beta\to\infty} \; \to \; 1
\end{equation}
and (see eq.(\ref{84}))
\begin{equation}
 \label{93}
C^{0}_{k^{\alpha}_{1}k^{\alpha}_{2}k^{\alpha}_{3}} \; \to \;
\frac{\Gamma(z^{\alpha}_{1}/\beta+ z^{\alpha}_{2}/\beta+ z^{\alpha}_{3}/\beta+1)}{
\Gamma(z^{\alpha}_{1}/\beta+1) \Gamma(z^{\alpha}_{2}/\beta+1) \Gamma(z^{\alpha}_{3}/\beta+1)}\Big|_{\beta\to\infty} \; \to \; 1
\end{equation}
Thus, after extracting the contributions with
$k^{\alpha}_{1} = k^{\alpha}_{2} = k^{\alpha}_{3} = 0$, the expression in eq.(\ref{88}) reduces to
\begin{eqnarray}
 \nonumber
&&P_{x,\epsilon}(f_{1}, f_{2}, f_{3}) \; = \;
\frac{T_{*}}{\zeta_{0}}
\frac{\partial^{3}}{\partial f_{1} \; \partial f_{2} \; \partial f_{3}}
\Bigl(\frac{\partial}{\partial f_{1}} + \frac{\partial}{\partial f_{2}} +\frac{\partial}{\partial f_{3}}\Bigr)
\sum_{n=1}^{\infty} \frac{(-1)^{n-1}}{n} \; \Bigl(\frac{\zeta_{0}}{T_{*}}\Bigr)^{n}
\times
\\
\nonumber
\\
\nonumber
&\times&
\Biggl[
\int_{0}^{+\infty} dy
\Biggl<
\Biggl(
\int_{{\cal C}}\frac{dz_{1}}{2\pi i z_{1}}
\exp\Bigl\{ z_{1}\bigl(\gamma x \xi + f_{1} - f_{01} - y \bigr)\Bigr\} \;
\int_{{\cal C}}\frac{dz_{2}}{2\pi i z_{2}}
\exp\Bigl\{ z_{2}\bigl(\gamma (x-\epsilon) \xi + f_{2} - f_{02} - y \bigr)\Bigr\}
\times
\\
\nonumber
\\
\label{94}
&\times&
\int_{{\cal C}}\frac{dz_{3}}{2\pi i z_{3}}
\exp\Bigl\{ z_{3}\bigl(\gamma \epsilon \xi + f_{3} - f_{03} - y \bigr)\Bigr\}
\; + \; 1
\Biggr)
\Biggr>_{\xi}
\Biggr]^{n}
\end{eqnarray}
Taking into account that
\begin{equation}
 \label{95}
\int_{{\cal C}}\frac{dz}{2\pi i z}
\exp\{ \lambda z\} \; = \; - \theta(-\lambda)
\end{equation}
we obtain
\begin{equation}
 \label{96}
P_{x,\epsilon}(f_{1}, f_{2}, f_{3}) \; = \;
\frac{T_{*}}{\zeta_{0}}
\frac{\partial^{3}}{\partial f_{1} \; \partial f_{2} \; \partial f_{3}}
\Bigl(\frac{\partial}{\partial f_{1}} + \frac{\partial}{\partial f_{2}} +\frac{\partial}{\partial f_{3}}\Bigr) \;
\ln\Bigl[ 1 \; + \; S(f_{1}, f_{2}, f_{3})\Bigr]
\end{equation}
or
\begin{equation}
 \label{97}
P_{x,\epsilon}(f_{1}, f_{2}, f_{3}) \; = \;
\frac{\partial^{3}}{\partial f_{1} \; \partial f_{2} \; \partial f_{3}}
\Biggl[
\Bigl(1 \; + \; S(f_{1}, f_{2}, f_{3})\Bigr)^{-1} \;
G(f_{1}, f_{2}, f_{3})
\Biggr]
\end{equation}
where
\begin{eqnarray}
\nonumber
S(f_{1}, f_{2}, f_{3}) &=&
\frac{\zeta_{0}}{T_{*}}
\int_{0}^{+\infty} dy
\int_{-\infty}^{+\infty} \frac{d\xi}{\sqrt{2\pi}} \exp\Bigl\{-\frac{1}{2}\xi^{2}\Bigr\} \;
\times
\\
\nonumber
\\
\label{98}
&\times&
\Biggl[
1 \; - \;
\theta\bigl(y + f_{01} - f_{1} - \gamma x \xi\bigr) \;
\theta\bigl(y + f_{02} - f_{2} - \gamma (x-\epsilon) \xi\bigr) \;
\theta\bigl(y + f_{03} - f_{3} - \gamma \epsilon \xi\bigr)
\Biggr]
\end{eqnarray}
and
\begin{eqnarray}
\nonumber
G(f_{1}, f_{2}, f_{3}) &=&
\frac{T_{*}}{\zeta_{0}} \;
\Bigl(\frac{\partial}{\partial f_{1}} + \frac{\partial}{\partial f_{2}} +\frac{\partial}{\partial f_{3}}\Bigr) \;
S(f_{1}, f_{2}, f_{3})
\\
\nonumber
\\
\label{99}
&=&
\int_{-\infty}^{+\infty} \frac{d\xi}{\sqrt{2\pi}} \exp\Bigl\{-\frac{1}{2}\xi^{2}\Bigr\}
\Biggl[
1 -
\theta\bigl(f_{01} - f_{1} - \gamma x \xi\bigr) \;
\theta\bigl(f_{02} - f_{2} - \gamma (x-\epsilon) \xi\bigr) \;
\theta\bigl(f_{03} - f_{3} - \gamma \epsilon \xi\bigr)
\Biggr]
\end{eqnarray}

\section{Two velocity probability density function}

In this Section using the general result for the directed polymers three-point free energy distribution function,
eqs.(\ref{97})-(\ref{99}), we are going to derive two velocity probability density function $P_{x}(v, v')$ of the
corresponding randomly forced Burgers problem. According to the discussion of Section II, eqs.(\ref{34})-(\ref{36}),
\begin{equation}
 \label{100}
P_{x}(v, v') \; = \; \lim_{\epsilon\to 0} \Biggl[\epsilon^{2} \int_{-\infty}^{+\infty} df_{2}
              P_{x,\epsilon}\bigl(f_{2}-\epsilon v,\,  f_{2},\, -\epsilon v'\bigr) \Biggr]
\end{equation}
where $v$ and $v'$ are two velocities at two spatial points separated by the distance $x$.
Explicitly, the expression for the function $P_{x,\epsilon}(f_{1}, f_{2}, f_{3})$, eq.(\ref{97}), reads
\begin{eqnarray}
 \nonumber
P_{x,\epsilon}(f_{1}, f_{2}, f_{3}) &=&
       \bigl(1 + S\bigr)^{-1} \, G_{123}''' \; - \;
\\
\nonumber
\\
\nonumber
&-&
       \bigl(1 + S\bigr)^{-2} \Bigl[S_{1}' G_{23}'' + S_{2}' G_{13}'' + S_{3}' G_{12}''
                                  + S_{12}'' G_{3}' + S_{13}'' G_{2}' + S_{23}'' G_{1}' + S_{123}''' G \Bigr] \; + \;
\\
\nonumber
\\
\nonumber
&+&
       2 \bigl(1 + S\bigr)^{-3} \Bigl[S_{1}' S_{2}' G_{3}' + S_{1}' S_{3}' G_{2}' + S_{2}' S_{3}' G_{1}'
                                  + \bigl(S_{1}' S_{23}'' + S_{2}' S_{13}'' + S_{3}' S_{12}''\bigr) \, G \Bigr] \; - \;
\\
\nonumber
\\
\label{101}
&-&
       6 \bigl(1 + S\bigr)^{-4} \, S_{1}'S_{2}' S_{3}' \; G
\end{eqnarray}
where we have introduced the notations $\Phi_{i}' \, \equiv \, \frac{\partial}{\partial f_{i}} \, \Phi$
and the functions $S = S(f_{1}, f_{2}, f_{3})$ and $G = G(f_{1}, f_{2}, f_{3})$ are given in eqs.(\ref{98}) and (\ref{99}).
Substituting this expression into eq.(\ref{100}) we find that the only non-zero contributions in the limit
$\epsilon\to 0$ come from two terms in the r.h.s. of eq.(\ref{101}): $\bigl(1 + S\bigr)^{-1} \, G_{123}'''$
and $-\bigl(1 + S\bigr)^{-2} S_{12}'' G_{3}'$ (both of which $\propto 1/\epsilon^{2}$) where (see Appendix A),
\begin{eqnarray}
 \label{102}
G_{123}'''\Big|_{\epsilon\to 0} &=& \frac{x}{\epsilon^{2} \gamma \sqrt{2\pi}}
               \exp\Bigl\{-\frac{1}{2\gamma^{2}} (v')^{2}\Bigr\}
              \delta\bigl(f_{2} - f_{0} + x \, v'\bigr) \;
              \delta\bigl(x\, v' - x \, v - 2f_{0}\bigr)
\\
\nonumber
\\
\label{103}
G_{3}'\Big|_{\epsilon\to 0} &=& \frac{1}{\epsilon \gamma \sqrt{2\pi}}
               \exp\Bigl\{-\frac{1}{2\gamma^{2}} (v')^{2}\Bigr\}
              \theta\bigl(f_{0} - f_{2} - x \, v'\bigr)
\\
\nonumber
\\
\label{104}
S_{12}''\Big|_{\epsilon\to 0} &=& -\frac{\zeta_{0}}{\epsilon \gamma T_{*} \sqrt{2\pi}}
               \exp\Bigl\{-\frac{1}{2\gamma^{2} x^{2}} \bigl(x\, v + 2f_{0})^{2}\Bigr\}
              \theta\bigl(x\, v + f_{0} + f_{2}\bigr)
\end{eqnarray}
Here, according to eqs.(\ref{71}), (\ref{62}), (\ref{56}) and (\ref{48}),
\begin{equation}
 \label{105}
f_{0} \; \equiv \; f_{01} \; = \; \frac{1}{2} (\beta m) \gamma^{2} x^{2} \; = \;
                                  \frac{1}{2} \sqrt{\zeta_{0}} \; T_{*}\cdot \frac{x^{2}}{R^{2}}
\end{equation}
\begin{equation}
 \label{106}
\gamma \; = \; \zeta_{0}^{-1/4} \frac{T_{*}}{R}
\end{equation}
and $T_{*}$ given in eq.(\ref{42}). Using explicit expression for $S(f_{1}, f_{2}, f_{3})$, eq.(\ref{98}), one finds
\begin{equation}
 \label{107}
\lim_{\epsilon\to 0} S\bigl(f_{2}-\epsilon v , f_{2}, -\epsilon v'\bigr) \; = \;
            \frac{\zeta_{0}}{\gamma T_{*} x} \int_{0}^{\infty} \frac{d\xi}{\sqrt{2\pi}}  \, \xi \;
             \exp\Bigl\{-\frac{1}{2\gamma^{2} x^{2}} \bigl(\xi + f_{0}- f_{2})^{2}\Bigr\}
\end{equation}
Substituting eqs.(\ref{101})-(\ref{107}) into eq.(\ref{100}) we get
\begin{eqnarray}
 \nonumber
P_{x}(v, v') &=&
     \int_{-\infty}^{+\infty} d f_{2}
\Biggl\{
 \frac{x}{\gamma\sqrt{2\pi}} \;
 \frac{\exp\Bigl\{-\frac{1}{2\gamma^{2}} (v')^{2}\Bigr\}
              \delta\bigl(f_{2} - f_{0} + x \, v'\bigr) \;
              \delta\bigl(x\, v' - x \, v - 2f_{0}\bigr)}{
      1 + \frac{\zeta_{0}}{\gamma T_{*} x} \int_{0}^{\infty} \frac{d\xi}{\sqrt{2\pi}}  \, \xi \;
             \exp\Bigl\{-\frac{1}{2\gamma^{2} x^{2}} \bigl(\xi + f_{0}- f_{2})^{2}\Bigr\}}
\; + \;
\\
\nonumber
\\
\nonumber
\\
\label{108}
&+&
\frac{\zeta_{0}}{2\pi \gamma^{2} T_{*}} \;
\frac{\exp\Bigl\{-\frac{1}{2\gamma^{2} x^{2}} \bigl(x\, v + 2f_{0})^{2} -\frac{1}{2\gamma^{2}} (v')^{2}   \Bigr\}
              \theta\bigl(x\, v + f_{0} + f_{2}\bigr) \theta\bigl(f_{0} - f_{2} - x \, v'\bigr) }{
   \Bigl[1 + \frac{\zeta_{0}}{\gamma T_{*} x} \int_{0}^{\infty} \frac{d\xi}{\sqrt{2\pi}}  \, \xi \;
             \exp\Bigl\{-\frac{1}{2\gamma^{2} x^{2}} \bigl(\xi + f_{0}- f_{2})^{2}\Bigr\}\Bigr]^{2}}
\Biggr\}
\end{eqnarray}
Introducing the notation (sf eq.(\ref{9a}))
\begin{equation}
\label{109}
v_{0} \; = \; \zeta_{0}^{3/4}\gamma \; = \; \sqrt{\zeta_{0}}\, \frac{T_{*}}{R}
                                    \; = \; \Bigl(\frac{\zeta_{0}^{3}}{2\pi}\Bigr)^{1/6}
                                            \Bigl(\frac{u}{R^{2}}\Bigr)^{1/3}
\end{equation}
and changing the integration variables, $f_{2} \to f_{0} - x v_{0} \eta$,  $\xi \to \gamma \, x \, \xi$
we eventually get the following result for the joint probability density function of two velocities
at the distance $x$:
\begin{equation}
\label{110}
P_{x}(v, v') \; = \; p_{0}\bigl(v, \, x\bigr) \, \delta\Bigl(v' - v - v_{0}\frac{x}{R}\Bigr) \; + \;
                 {\cal P}_{x}\bigl(v, \, v'\bigr)\, \theta\Bigl(v + v_{0}\frac{x}{R} - v'\Bigr)
\end{equation}
where
\begin{equation}
\label{111}
p_{0}(v, x) \; = \; \frac{\zeta_{0}^{3/4}}{v_{0} \sqrt{2\pi}} \;
       \frac{\exp\Bigl\{-\frac{1}{2}\zeta_{0}^{3/2} \Bigl(\frac{v}{v_{0}} + \frac{x}{R}\Bigr)^{2}\Bigr\}}{
      \Bigl[
      1 + \zeta_{0}^{3/4}\frac{x}{R} \int_{0}^{\infty} \frac{d\xi}{\sqrt{2\pi}}  \; \xi \;
             \exp\Bigl\{
             -\frac{1}{2} \Bigl[\xi + \zeta_{0}^{3/4}\Bigl(\frac{v}{v_{0}} + \frac{x}{R}\Bigr)\Bigr]^{2}
             \Bigr\}
      \Bigr]   }
\end{equation}
and
\begin{equation}
\label{112}
{\cal P}_{x}(v, v') \; = \;
  \frac{\zeta_{0}^{3} \, x}{2\pi \, v_{0}^{2} \, R}
  \int_{v'/v_{0}}^{v/v_{0} + x/R}d\eta \,
  \frac{\exp\Bigl\{ -\frac{1}{2}\zeta_{0}^{3/2}
     \Bigl[ \Bigl(\frac{v}{v_{0}} + \frac{x}{R}\Bigr)^{2} +\Bigl(\frac{v'}{v_{0}}\Bigr)^{2}\Bigr] \Bigr\}}{
     \Bigl[ 1 + \zeta_{0}^{3/4}\frac{x}{R} \int_{0}^{\infty} \frac{d\xi}{\sqrt{2\pi}}  \; \xi \;
     \exp\Bigl\{-\frac{1}{2} \bigl(\xi + \zeta_{0}^{3/4}\eta \bigr)^{2} \Bigr\}
      \Bigr]^{2}   }
\end{equation}

\section{Probability distribution function of the velocity difference}

Using the joint distribution function $P_{x}(v, \, v')$ of two velocities $v$ and $v'$ at distance $x$ derived above,
eqs.(\ref{110})-(\ref{112}), the probability density function of the velocity difference $w = v' - v$ can be obtained as follows
\begin{equation}
\label{113}
P_{x}(w) \; = \; \int_{-\infty}^{+\infty} dv \; P_{x}(v,\; v + w)
\end{equation}
Substituting here eqs.(\ref{110})-(\ref{112}) we get
\begin{equation}
\label{114}
P_{x}(w) \; = \; p_{0}(x) \, \delta\Bigl(w - v_{0}\frac{x}{R}\Bigr) \; + \;
                 {\cal P}_{x}(w)\, \theta\Bigl(v_{0}\frac{x}{R} - w\Bigr)
\end{equation}
where
\begin{equation}
\label{115}
p_{0}(x) \; = \; \frac{\zeta_{0}^{3/4}}{v_{0} \sqrt{2\pi}} \; \int_{-\infty}^{+\infty} dv
       \frac{\exp\Bigl\{-\frac{1}{2}\zeta_{0}^{3/2} \Bigl(\frac{v}{v_{0}} + \frac{x}{R}\Bigr)^{2}\Bigr\}}{
      \Bigl[
      1 + \zeta_{0}^{3/4}\frac{x}{R} \int_{0}^{\infty} \frac{d\xi}{\sqrt{2\pi}}  \; \xi \;
             \exp\Bigl\{
             -\frac{1}{2} \Bigl[\xi + \zeta_{0}^{3/4}\Bigl(\frac{v}{v_{0}} + \frac{x}{R}\Bigr)\Bigr]^{2}
             \Bigr\}
      \Bigr]   }
\end{equation}
and
\begin{equation}
\label{116}
{\cal P}_{x}(w) \; = \;
  \frac{\zeta_{0}^{3} \, x}{2\pi \, v_{0}^{2} \, R} \; \int_{-\infty}^{+\infty} dv
  \int_{v'/v_{0}}^{v/v_{0} + x/R}d\eta \,
  \frac{\exp\Bigl\{ -\frac{1}{2}\zeta_{0}^{3/2}
     \Bigl[ \Bigl(\frac{v}{v_{0}} + \frac{x}{R}\Bigr)^{2} +\Bigl(\frac{v + w}{v_{0}}\Bigr)^{2}\Bigr] \Bigr\}}{
     \Bigl[ 1 + \zeta_{0}^{3/4}\frac{x}{R} \int_{0}^{\infty} \frac{d\xi}{\sqrt{2\pi}}  \; \xi \;
     \exp\Bigl\{-\frac{1}{2} \bigl(\xi + \zeta_{0}^{3/4}\eta \bigr)^{2} \Bigr\}
      \Bigr]^{2}   }
\end{equation}
Changing the integration variables: $v =  -\frac{x}{R} v_{0} + \zeta_{0}^{-3/4} v_{0} \, s$ and
$\eta  =  (v+w)/v_{0} + \zeta_{0}^{-3/4}  z$, and introducing rescaled (dimensionless)
distance
\begin{equation}
\label{117}
r \; \equiv \; \zeta_{0}^{3/4} \frac{x}{R}
\end{equation}
and rescaled (dimensionless) velocity difference
\begin{equation}
\label{118}
\omega \; \equiv \; \zeta_{0}^{3/4} \frac{w}{v_{0}} \; = \; \zeta_{0}^{3/4} \frac{(v' - v)}{v_{0}}
\end{equation}
for the corresponding probability density function $P_{r}(\omega)$
we obtain the following final result:
\begin{equation}
\label{119}
P_{r}(\omega) \; = \; p_{0}(r) \, \delta\bigl(\omega - r\bigr) \; + \;
                 {\cal P}_{r}(\omega)\, \theta\bigl(r - \omega\bigr)
\end{equation}
where
\begin{equation}
\label{120}
p_{0}(r) \; = \; \int_{-\infty}^{+\infty} \frac{ds}{\sqrt{2\pi}}
       \frac{\exp\Bigl\{-\frac{1}{2}s^{2}\Bigr\}}{
      \Bigl[1 + r \int_{0}^{\infty} \frac{d\xi}{\sqrt{2\pi}}  \; \xi \;
             \exp\Bigl\{-\frac{1}{2} \bigl(\xi + s\bigr)^{2}\Bigr\}
      \Bigr]}
\end{equation}
and
\begin{equation}
\label{121}
{\cal P}_{r}(\omega) \; = \;
  r\, \int_{-\infty}^{+\infty} \frac{ds}{\sqrt{2\pi}} \int_{0}^{r-\omega} \frac{dz}{\sqrt{2\pi}} \,
  \frac{\exp\Bigl\{ -\frac{1}{2} s^{2} - \frac{1}{2} \bigl(s + \omega -r\bigr)^{2}\Bigr\}}{
     \Bigl[ 1 + r \int_{0}^{\infty} \frac{d\xi}{\sqrt{2\pi}}  \; \xi \;
     \exp\Bigl\{-\frac{1}{2} \bigl(\xi + z + s + \omega -r\bigr)^{2} \Bigr\}
      \Bigr]^{2}   }
\end{equation}
It is evident that this function is positively defined and it can be easily checked that for any value of $r$ it is normalized:
\begin{equation}
 \label{122}
\int_{-\infty}^{+\infty} d\omega \, P_{r}(\omega) \, = \, 1
\end{equation}
\begin{figure}[h]
\begin{center}
   \includegraphics[width=13.0cm]{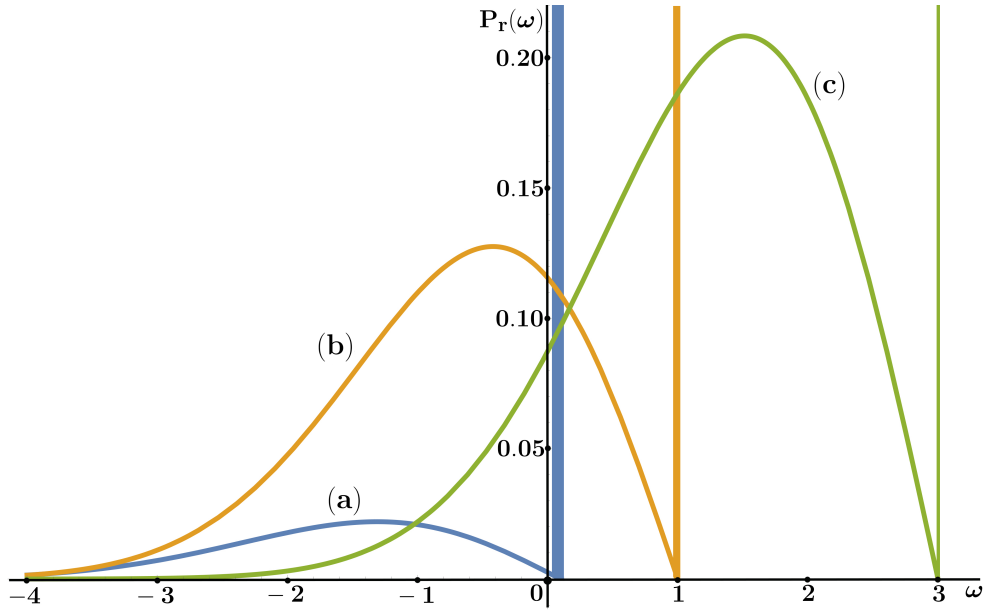}
\caption[]{Probability density function $P_{r}(\omega)$, eqs.(\ref{119})-(\ref{121}) for:
           (a) $r = 0.1$; (b) $r=1$; (c) $r=3$. The vertical lines at $\omega = r$ represent the $\delta$-functions, 
          and the difference in the  thickness of these lines symbolizes the relative values 
          of the corresponding weights $p_{0}(r)$, eq.(\ref{120}), 
          which decrease with increasing $r$.}
\end{center}
\label{figure4}
\end{figure}
We see that the distribution function $P_{r}(\omega)$ has rather specific structure (see Fig.4).
According to eq.(\ref{119}) for a given (rescaled) distance $r$ possible values of the (rescaled) velocity
difference $\omega$ are bounded from above: $\omega \leq r$, or in terms of the original values, $(v'-v) \leq \frac{x}{R} v_{0}$
In other words, at a given distance $x$
between two points at which we measure two velocities $v$ and $v'$, their difference can not be bigger then
$v_{0} \, x/R$, where, according to eq.(\ref{109}), $v_{0} \propto \bigl(u/R^{2}\bigr)^{1/3}$ is the typical flow velocity
at the injection scale $R$ of the random force of the strength $u$.
Moreover, at $\omega = r$
(or at $(v'-v) = \frac{x}{R} v_{0}$) the distribution function exhibits the $\delta$-function singularity.

Let us investigate the statistical properties of the velocity difference at small distances, $x \ll R$, or $r \ll 1$.
In the limit of small values of the parameter $r$, the probability density function $P_{r}(\omega)$, eqs.(\ref{119})-(\ref{121}),
takes much more simple form:
\begin{equation}
 \label{123}
P_{r}(\omega) \; \simeq \; \Bigl(1 - \frac{1}{\sqrt{\pi}} \, r\Bigr) \, \delta\bigl(\omega - r\bigr) \; + \;
         \frac{1}{\sqrt{\pi}} \, r \, \bigl(r - \omega\bigr) \, \exp\Bigl\{-\frac{1}{2} (r - \omega)^{2}\Bigr\} \,
         \theta\bigl(r - \omega\bigr)
\end{equation}
For even moments of the velocity difference $\langle \omega^{2n}\rangle$ we find:
\begin{equation}
\label{124}
\langle \omega^{2n}\rangle \; = \; \int_{-\infty}^{+\infty} d\omega \; \omega^{2n} \; P_{r}(\omega) \; \simeq \;
     r^{2n} \; + \; C(n) \, r 
\end{equation}
where $C(n) \; = \; 2^{2n+1} \, \Gamma(1+n)$. 
Then, the analytic continuation of the above result for arbitrary real values
of the parameter $2n \to q$, in the limit $r \ll 1$ yields:
\begin{equation}
\label{124a}
\langle \omega^{q}\rangle \; \simeq \;
     r^{q} \; + \; C(q/2) \, r \; \simeq \;
     \left\{
        \begin{array}{ll}
           r^{q} \; , \; \; \mbox{for} \; q \; \leq \; 1 \, ;
       \\
       \\
           C(q/2) \, r \; , \; \; \mbox{for} \; q \; > \; 1 \, .
        \end{array}
\right.
\end{equation}
Finally, introducing the exponent $\zeta(q)$ according to the definition
$\langle \omega^{q}\rangle = r^{\zeta(q)}$  we recover the typical strong
intermittency behavior \cite{Bouch-Mez-Par}(see Fig.1):
\begin{equation}
 \label{125}
\zeta(q) \; \simeq \;
\left\{
        \begin{array}{ll}
           q \; , \; \; \mbox{for} \; q \; \leq \; 1 \, ;
       \\
       \\
           1 \; , \; \; \mbox{for} \; q \; > \; 1 \, .
        \end{array}
\right.
\end{equation}

\section{Conclusions}

In this paper we studied the statistical properties of of the velocity field $v(x,t)$
in  the one-dimensional randomly forced Burgers turbulence, eq.(\ref{1}). This system is known to be equivalent to
the model of directed polymers in a random potential, eqs.(\ref{4})-(\ref{7}),
such that the the viscosity parameter $\nu$ in the Burger's equation
is proportional to the temperature in the directed polymer system, $\nu = \frac{1}{2} T$, and the velocity $v(x,t)$
in the Burger's equation is the negative spatial derivative of the free energy $F(x,t)$ of the directed polymers.
The parameter which characterizes the level
of turbulence of the velocity field in the Burgers problem is the Reynolds number $Re$
which in terms of the directed polymers notations is expressed as $Re = 2\bigl(u R\bigr)^{1/3}/T$
where $R$ and $u$ are the correlation length and the strength of the random potential, eqs.(\ref{5})-(\ref{6}).
Thus the strong turbulence regime where $Re \to \infty$  corresponds to the zero-temperature limit
in the directed polymers system. In this limit in terms of the replica technique
a general expression for the joint distribution function of two velocities $v(-x/2,t)$ and $v(x/2,t)$
separated by a finite distance $x$ has been derived, eqs.(\ref{110})-(\ref{112}).
Besides we have obtained an explicit expression for the probability density function for the
corresponding velocity increment $w = v(-x/2,t)-v(x/2,t)$, eqs.(\ref{119})-(\ref{121}), which
was shown to exhibit rather specific structure. Namely, for any given distance $x$ the values of the velocity increment
$w$ are bounded from above:  $w \leq \frac{x}{R} v_{0}$, where  $v_{0} \propto \bigl(u/R^{2}\bigr)^{1/3}$
is the typical flow velocity at the injection scale $R$ of the random potential.
Moreover, at $w = \frac{x}{R} v_{0}$ the distribution function exhibits the $\delta$-function singularity
which means that at a given distance $x$ the difference of two velocities $w = v(-x/2,t)-v(x/2,t)$ has a {\it finite}
probability  to be equal to $\frac{x}{R} v_{0}$. Using this distribution function at length scales
much smaller than the injection length of the random potential, $x \ll R$ we have computed the
moments of the velocity increment $\langle \omega^{q}\rangle$, eq.(\ref{124a}).
Introducing the exponent $\zeta(q)$ according to the definition
$\langle \omega^{q}\rangle \; \simeq \; r^{\zeta(q)}$
we have demonstrated that the function $\zeta(q)$ exhibits the behavior
typical for strong intermittency phenomena, eq.(\ref{125}), Fig.1.

Finally a few remarks about the status of the obtained results. First of all, as the considerations
has been performed in the framework of the heuristic replica method, the proposed derivation 
can be considered as rigorous. Moreover, at the moment it is also difficult to say whether the obtained results
are exact or not: on one hand, no approximations have been used in the performed calculations, but on the other hand,
the considered derivation is based on the unproved crucial  assumption about
the vector replica symmetry breaking structure of the $N$-particle bosonic wave function in the zero-temperature limit
which, in particular, contains undefined numerical factor $\zeta_{0}$, eq.(\ref{56})
(\cite{zero-T}, Section III). All that means that further more systematic study of the considered problem.
is required.

\acknowledgments

I am grateful to Kostya Khanin for numerous useful discussions.

I would like to thank the mathematical research institute MATRIX in Australia where part of this research was performed.

\vspace{15mm}

\begin{center}

\appendix{\large \bf Appendix A}

\end{center}

\newcounter{A}
\setcounter{equation}{0}
\renewcommand{\theequation}{A.\arabic{equation}}

\vspace{5mm}

In this technical Appendix the derivatives of the functions $S(f_{1},f_{2},f_{3})$ and $G(f_{1},f_{2},f_{3})$,
eqs.(\ref{98}) and (\ref{99}), will be calculated in the limit $\epsilon\to 0$.
Redefining:
\begin{eqnarray}
 \label{A1}
\xi &=& \frac{1}{\gamma x}\tilde{\xi}
\\
\label{A2}
\epsilon &=& \tilde{\epsilon} \, x
\end{eqnarray}
and introducing velocities $v$ and $v'$ instead of
$f_{1}$ and $f_{3}$,
\begin{eqnarray}
 \label{A3}
f_{1} &=& -\tilde{\epsilon} \, x\, v \; + \; f_{2}
\\
\label{A4}
f_{3} &=& -\tilde{\epsilon} \, x\, v'
\end{eqnarray}
we get
\begin{eqnarray}
 \label{A5}
\frac{\partial}{\partial f_{1}} &=& - \frac{1}{\tilde{\epsilon} \, x} \, \frac{\partial}{\partial v}
\\
\label{A6}
\frac{\partial}{\partial f_{3}} &=& - \frac{1}{\tilde{\epsilon} \, x} \, \frac{\partial}{\partial v'}
\end{eqnarray}
According to eqs.(\ref{71}), (\ref{62}), (\ref{56}) and (\ref{48}), in the first order in $\epsilon \to 0$ we have:
\begin{eqnarray}
 \label{A7}
f_{01} &=& \frac{1}{2} (\beta m) \gamma^{2} x^{2} \; = \; \frac{1}{2} \sqrt{\zeta_{0}} \, T_{*} \Bigl(\frac{x}{R}\Bigr)^{2} \equiv f_{0}
\\
\label{A8}
f_{01} &=& \frac{1}{2} (\beta m) \gamma^{2} (x-\epsilon)^{2} \; \simeq \; f_{0} \; + \; 2 \tilde{\epsilon} f_{0}
\\
\label{A9}
f_{03} &=& \frac{1}{2} (\beta m) \gamma^{2} \epsilon^{2} \; \to \; 0
\end{eqnarray}
Substituting eqs.(\ref{A1})-(\ref{A9}) into eqs.(\ref{98}) and (\ref{99}) we find
\begin{eqnarray}
\nonumber
S(v, f_{2}, v') &=&
\frac{\zeta_{0}}{T_{*}\gamma x}
\int_{0}^{+\infty} dy
\int_{-\infty}^{+\infty} \frac{d\tilde{\xi}}{\sqrt{2\pi}} \exp\Bigl\{-\frac{\tilde{\xi}^{2}}{2\gamma^{2} x^{2}}\Bigr\} \;
\times
\\
\nonumber
\\
\label{A10}
&\times&
\Biggl[
1 \; - \;
\theta\bigl(y + f_{0} - f_{2} + \tilde{\epsilon} x v - \tilde{\xi}\bigr) \;
\theta\bigl(y + f_{0} - f_{2} - 2\tilde{\epsilon}f_{0} + \tilde{\epsilon}\tilde{\xi} - \tilde{\xi}\bigr) \;
\theta\bigl(y + \tilde{\epsilon} x v'  - \tilde{\epsilon}\tilde{\xi}\bigr)
\Biggr]
\end{eqnarray}
and
\begin{equation}
\label{A11}
G(v, f_{2}, v') = \frac{1}{\gamma x}
\int_{-\infty}^{+\infty} \frac{d\tilde{\xi}}{\sqrt{2\pi}} \exp\Bigl\{-\frac{\tilde{\xi}^{2}}{2\gamma^{2} x^{2}}\Bigr\}
\Biggl[
1 - \theta\bigl(f_{0} - f_{2} + \tilde{\epsilon} x v - \tilde{\xi}\bigr) \;
    \theta\bigl(f_{0} - f_{2} - 2\tilde{\epsilon}f_{0} + \tilde{\epsilon}\tilde{\xi} - \tilde{\xi}\bigr) \;
    \theta\bigl(\tilde{\epsilon} x v' - \tilde{\epsilon}\tilde{\xi}\bigr)
\Biggr]
\end{equation}
The calculation of the derivatives of these functions is straightforward. For example, for the derivative
$G'_{1} \equiv -\frac{1}{\tilde{\epsilon} x} \, \frac{\partial}{\partial v} G$ we get
\begin{equation}
 \label{A12}
G'_{1} \; = \; \frac{1}{\gamma x}
\int_{-\infty}^{+\infty} \frac{d\tilde{\xi}}{\sqrt{2\pi}} \exp\Bigl\{-\frac{\tilde{\xi}^{2}}{2\gamma^{2} x^{2}}\Bigr\}
    \delta\bigl(f_{0} - f_{2} + \tilde{\epsilon} x v - \tilde{\xi}\bigr) \;
    \theta\bigl(f_{0} - f_{2} - 2\tilde{\epsilon}f_{0} + \tilde{\epsilon}\tilde{\xi} - \tilde{\xi}\bigr) \;
    \theta\bigl(\tilde{\epsilon} x v' - \tilde{\epsilon}\tilde{\xi}\bigr)
\end{equation}
Taking the limit $\epsilon\to 0$ we find:
\begin{equation}
 \label{A13}
\lim_{\epsilon\to 0} G'_{1} \; = \; \frac{1}{\gamma x\sqrt{2\pi}}
    \exp\Bigl\{-\frac{(f_{2}-f_{0})^{2}}{2\gamma^{2} x^{2}}\Bigr\}
    \theta\bigl(f_{2} - f_{0} + x v'\bigr) \;
    \theta\bigl(-f_{0} - f_{2} - x v\bigr)
\end{equation}
In a similar way we obtain the rest of the derivatives:
\begin{eqnarray}
 \label{A14}
\lim_{\epsilon\to 0} G'_{2}  &=& \frac{1}{\gamma x\sqrt{2\pi}}
    \exp\Bigl\{-\frac{(f_{2}-f_{0})^{2}}{2\gamma^{2} x^{2}}\Bigr\} \;
    \theta\bigl(f_{2} - f_{0} + x v'\bigr) \;
    \theta\bigl(f_{0} + f_{2} + x v\bigr)
\\
\nonumber
\\
\label{A15}
\lim_{\epsilon\to 0} G'_{3}  &=& \frac{1}{\gamma \epsilon\sqrt{2\pi}}
    \exp\Bigl\{-\frac{(v')^{2}}{2\gamma^{2} x^{2}}\Bigr\} \;
    \theta\bigl(f_{0} - f_{2} - x v'\bigr)
\\
\nonumber
\\
 \label{A16}
\lim_{\epsilon\to 0} G''_{12}  &=& -\frac{1}{\gamma \epsilon\sqrt{2\pi}}
    \exp\Bigl\{-\frac{(x\, v + 2 f_{0})^{2}}{2\gamma^{2} x^{2}}\Bigr\} \;
    \theta\bigl(x v'- xv - 2f_{0}\bigr) \;
    \delta\bigl(f_{2} + f_{0} + x v\bigr)
\\
\nonumber
\\
\label{A17}
\lim_{\epsilon\to 0} G''_{13}  &=& -\frac{1}{\gamma \epsilon\sqrt{2\pi}}
    \exp\Bigl\{-\frac{(v')^{2}}{2\gamma^{2} x^{2}}\Bigr\} \;
    \theta\bigl(x v'- xv - 2f_{0}\bigr) \;
    \delta\bigl(f_{2} - f_{0} + x v'\bigr)
\\
\nonumber
\\
\label{A18}
\lim_{\epsilon\to 0} G''_{13}  &=& -\frac{1}{\gamma \epsilon\sqrt{2\pi}}
    \exp\Bigl\{-\frac{(v')^{2}}{2\gamma^{2} x^{2}}\Bigr\} \;
    \theta\bigl(x v - xv' + 2f_{0}\bigr) \;
    \delta\bigl(f_{2} - f_{0} + x v'\bigr)
\\
\nonumber
\\
\label{A19}
\lim_{\epsilon\to 0} G'''_{123}  &=& \frac{1}{\gamma \epsilon^{2}\sqrt{2\pi}}
    \exp\Bigl\{-\frac{(v')^{2}}{2\gamma^{2} x^{2}}\Bigr\} \;
    \delta\bigl(x v' - xv + 2f_{0}\bigr) \;
    \delta\bigl(f_{2} - f_{0} + x v'\bigr)
\end{eqnarray}
\begin{eqnarray}
 \label{A20}
\lim_{\epsilon\to 0} S'_{1}  &=& \frac{\zeta_{0}}{T_{*}\gamma x\sqrt{2\pi}}
    \int_{0}^{\infty} dy \;
    \exp\Bigl\{-\frac{(y - f_{2} + f_{0})^{2}}{2\gamma^{2} x^{2}}\Bigr\} \;
    \theta\bigl(y - f_{2} - f_{0} - x v\bigr)
\\
\nonumber
\\
 \label{A21}
\lim_{\epsilon\to 0} S'_{2}  &=& \frac{\zeta_{0}}{T_{*}\gamma x\sqrt{2\pi}}
    \int_{0}^{\infty} dy \;
    \exp\Bigl\{-\frac{(y - f_{2} + f_{0})^{2}}{2\gamma^{2} x^{2}}\Bigr\} \;
    \theta\bigl(f_{2} + f_{0} + x v -y\bigr)
\\
\nonumber
\\
 \label{A22}
\lim_{\epsilon\to 0} S'_{3}  &=& \frac{\zeta_{0}}{T_{*}\gamma x\sqrt{2\pi}}
    \int_{0}^{\infty} dy \;
    \exp\Bigl\{-\frac{(y + x v')^{2}}{2\gamma^{2} x^{2}}\Bigr\} \;
    \theta\bigl(f_{0} - f_{2} - x v' -y\bigr)
    \theta\bigl(f_{0} - f_{2} - x v'\bigr)
\\
\nonumber
\\
 \label{A23}
\lim_{\epsilon\to 0} S''_{12}  &=& -\frac{\zeta_{0}}{T_{*}\gamma \epsilon\sqrt{2\pi}}
    \exp\Bigl\{-\frac{(xv + 2f_{0})^{2}}{2\gamma^{2} x^{2}}\Bigr\} \;
    \theta\bigl(xv + f_{0} + f_{2}\bigr)
\\
\nonumber
\\
 \label{A24}
\lim_{\epsilon\to 0} S''_{13}  &=& -\frac{\zeta_{0}}{T_{*}\gamma x\sqrt{2\pi}}
    \exp\Bigl\{-\frac{(f_{2} - f_{0})^{2}}{2\gamma^{2} x^{2}}\Bigr\} \;
    \theta\bigl(-f_{2} - f_{0} - x v\bigr)
    \theta\bigl(f_{0} - f_{2} - x v'\bigr)
\\
\nonumber
\\
 \label{A25}
\lim_{\epsilon\to 0} S''_{23}  &=& -\frac{\zeta_{0}}{T_{*}\gamma x\sqrt{2\pi}}
    \exp\Bigl\{-\frac{(f_{2} - f_{0})^{2}}{2\gamma^{2} x^{2}}\Bigr\} \;
    \theta\bigl(f_{2} + f_{0} + x v\bigr)
    \theta\bigl(f_{0} - f_{2} - x v'\bigr)
\\
\nonumber
\\
 \label{A26}
\lim_{\epsilon\to 0} S'''_{123}  &=& \frac{\zeta_{0}}{T_{*}\gamma \epsilon\sqrt{2\pi}}
    \exp\Bigl\{-\frac{(xv + 2f_{0})^{2}}{2\gamma^{2} x^{2}}\Bigr\} \;
    \theta\bigl(xv + 2f_{0} - xv'\bigr)
\end{eqnarray}
We see that in the limit $\epsilon\to 0$ according to the above expressions, eqs.(\ref{A13})-(\ref{A26}),
the only non-zero contribution to the function $P_{x}(v, v')$ in eqs.(\ref{100})-(\ref{101}) are given
by the two terms: $G'''_{123} \propto 1/\epsilon^{2}$, eq.(\ref{A19}),  and the product
$G'_{3} \, S''_{12} \propto 1/\epsilon^{2}$, eqs.(\ref{A15}) and (\ref{A23}).



\end{document}